\documentclass[aip,pop,reprint,numerical,twocolumn]{revtex4-1}

\usepackage{commands}
\DeclareMathOperator*{\argmax}{arg\,max}
\usepackage{circuitikz}
\graphicspath{{Figures/}} %Setting the graphicspath

\begin{document}

\title{Exploring the parameter space of MagLIF implosions using similarity scaling.~~II.~Current scaling}
\date{\today}
\author{D.~E.~Ruiz}
\email{deruiz@sandia.gov}
\affiliation{Sandia National Laboratories, P.O. Box 5800, Albuquerque, NM 87185, USA}
\author{P.~F.~Schmit}
\affiliation{Lawrence Livermore National Laboratory, Livermore, CA 94550, USA}
\author{D.~A.~Yager-Elorriaga}
\affiliation{Sandia National Laboratories, P.O. Box 5800, Albuquerque, NM 87185, USA}
\author{M.~R.~Gomez}
\affiliation{Sandia National Laboratories, P.O. Box 5800, Albuquerque, NM 87185, USA}
\author{M.~R.~Weis}
\affiliation{Sandia National Laboratories, P.O. Box 5800, Albuquerque, NM 87185, USA}
\author{C.~A.~Jennings}
\affiliation{Sandia National Laboratories, P.O. Box 5800, Albuquerque, NM 87185, USA}
\author{A.~J.~Harvey-Thompson}
\affiliation{Sandia National Laboratories, P.O. Box 5800, Albuquerque, NM 87185, USA}
\author{P.~F.~Knapp}
\affiliation{Sandia National Laboratories, P.O. Box 5800, Albuquerque, NM 87185, USA}
\author{S.~A.~Slutz}
\affiliation{Sandia National Laboratories, P.O. Box 5800, Albuquerque, NM 87185, USA}
\author{D.~J.~Ampleford}
\affiliation{Sandia National Laboratories, P.O. Box 5800, Albuquerque, NM 87185, USA}
\author{K.~Beckwith}
\affiliation{Sandia National Laboratories, P.O. Box 5800, Albuquerque, NM 87185, USA}
\author{M.~K.~Matzen}
\affiliation{Sandia National Laboratories, P.O. Box 5800, Albuquerque, NM 87185, USA}

%%%%%%%%%%%%%%%%%%%%%%%%%%%%%%%%%%%%%%%%%%%%%%%%%
%%%%%%%%%%%%%%%%%%%%%%%%%%%%%%%%%%%%%%%%%%%%%%%%%
%%%%%%%%%%%%%%%%%%%%%%%%%%%%%%%%%%%%%%%%%%%%%%%%%

\begin{abstract}

Magnetized Liner Inertial Fusion (MagLIF) is a magneto-inertial-fusion (MIF) concept, which is presently being studied on the Z Pulsed Power Facility.  The MagLIF platform has achieved interesting plasma conditions at stagnation and produced significant fusion yields in the laboratory.  Given the relative success of MagLIF, there is a strong interest to scale the platform to higher peak currents. However, scaling MagLIF is not entirely straightforward due to the large dimensionality of the experimental input parameter space and the large number of distinct physical processes involved in MIF implosions.  In this work, we propose a novel method to scale MagLIF loads to higher currents.  Our method is based on similarity (or similitude) scaling and attempts to preserve much of the physics regimes already known or being studied on today's Z pulsed-power driver.  By avoiding significant deviations into unexplored and/or less well-understood regimes, the risk of unexpected outcomes on future scaled-up experiments is reduced.  Using arguments based on similarity scaling, we derive the scaling rules for the experimental input parameters characterizing a MagLIF load (as functions of the characteristic current driving the implosion).  We then test the estimated scaling laws for various metrics measuring performance against results of 2D radiation--magneto-hydrodynamic \textsc{hydra} simulations.  Agreement is found between the scaling theory and the simulation results.

\end{abstract}

\maketitle

%%%%%%%%%%%%%%%%%%%%%%%%%%%%%%%%%%%%%%%%%%%%%%%%%
%%%%%%%%%%%%%%%%%%%%%%%%%%%%%%%%%%%%%%%%%%%%%%%%%
%%%%%%%%%%%%%%%%%%%%%%%%%%%%%%%%%%%%%%%%%%%%%%%%%
\section{Introduction} 

Magneto-inertial fusion (MIF) is one of the main approaches in inertial-confinement fusion (ICF).  Traditional ICF approaches based on laser-driven implosions require high implosion velocities to achieve plasma conditions at stagnation that can produce significant fusion yields in the laboratory.  MIF introduces strong magnetic fields in the fuel in order to relax the stringent requirements {\color{black}on the high implosion velocities and high pressures (both for the external drive and at stagnation).\cite{Widner_1977,Lindemuth_1981,Lindemuth:1983aa,Lindemuth:2015fu}}  One particular MIF concept is the Magnetized Liner Inertial Fusion (MagLIF) platform,\cite{Slutz:2010hd} which is currently being studied at the Z Pulsed Power facility at Sandia National Laboratories.\cite{Gomez:2014eta,Knapp:2019gf,Gomez:2019bg,Gomez:2020cd,YagerElorriaga:2022cp,Sinars:2020bv}  The Z facility delivers a 20-MA electrical current pulse to the cylindrical MagLIF z-pinch, which then implodes under the action of the Lorentz force.\citep{Gomez:2020cd}  Since MagLIF utilizes a relatively thick and heavy metallic cylindrical tamper, or liner, the achievable implosion velocities are substantially lower than those achieved in traditional ICF.  Therefore, the fuel is not shock-heated; instead, a 2--4-kJ 1-TW laser is used to preheat the fuel in order to {\color{black}increase the initial fuel adiabat}.\cite{HarveyThompson:2018dd,HarveyThompson:2019ff,Weis:2021id}  The implosions are considerably slower (on the order of 100 ns) so the fuel must be premagnetized to reduce thermal conduction losses.  This is achieved by external electromagnetic coils which provide a 10--16 T axial magnetic field.  The combination of these key elements has led to significant thermonuclear yield production in laboratory experiments\cite{Gomez:2014eta,Knapp:2019gf,Gomez:2019bg,Gomez:2020cd,YagerElorriaga:2022cp} and plasma magnetization inferred via secondary DT neutron emission.\cite{Schmit:2014fg,Knapp:2015kc,Lewis:2021kz}

%{\color{black}The implosions are considerably slower (on the order of 100 ns) than those in laser-driven ICF, so the fuel must be axially premagnetized to mitigate thermal-conduction losses.  The fuel is also not appreciably shock heated by the shock traversing the liner material, so the fuel is preheated instead by a 2--4-kJ, 1-TW laser to increase the fuel adiabat.}\cite{HarveyThompson:2019ff,HarveyThompson:2018dd,Weis:2021id} \\

Given the relative success of MagLIF and its demonstrated confinement parameter $P\tau\simeq 3.6$~Gbar-ns at $20$-MA peak current,\cite{foot:Knapp} there is high interest in scaling the platform to higher peak currents, \eg, to 45 MA or even 60 MA.  However, scaling MagLIF is not straightforward.  The space of experimental input parameters describing MagLIF is at least eight dimensional.  Aside from peak current, experimental parameters include the current rise time, the liner inner and outer radii, the liner material, the height of the liner, the delivered preheat energy, the imposed external magnetic field, and the initial fuel density.  Given the need to explore a relatively large parameter space, scoping future MagLIF designs at higher peak currents with radiation--magneto-hydrodynamic (rad-MHD) modeling tools can become expensive in terms of the computational resources.

Nevertheless, several numerical studies have explored the potential for MagLIF to generate high fusion yields on future, higher-energy pulsed-power drivers.\cite{Slutz:2012gp,Sefkow:2014ik,McBride:2015ga,Slutz:2016cf,Slutz:2018iq}  In order to reduce the dimensionality of the design space, these studies often constrain certain design parameters such as the current rise-time of the pulsed-power generator, the liner height, the liner aspect ratio, the liner material (usually beryllium or gold in some cases\citep{Slutz:2018iq}), and the external magnetic field.  Then, with the remaining basic experimental input parameters (liner outer radius, fuel preheat, and initial fuel density), an optimized configuration is sought that maximizes the fusion yield or energy gain of the implosions at a given peak current.\cite{Slutz:2016cf,Slutz:2018iq}  Results from these \emph{optimized-scaling} studies were obtained from thousands of 1D \textsc{lasnex} simulations and subsequent more refined 2D \textsc{lasnex} simulations.  Interestingly, these optimized-scaling studies predict $Y\simeq18$-MJ and $Y\simeq 440$-MJ DT yields for scaled ``gas-burning" MagLIF platforms, \ie loads with gaseous fuel configurations, driven at $I_{\rm max} \simeq 48$-MA and $I_{\rm max} \simeq65$-MA peak currents, respectively.\cite{Slutz:2016cf}

One disadvantage of the optimized-scaling approach is that the solution to the optimization problem may have implosion dynamics and energy-transport regimes different  from those presently studied on the Z facility.  These changes evidently increment the risk of achieving the desired performance of extrapolated MagLIF loads even when the rad-MHD modeling tools may account for these changes.  Following the results of \Refa{foot:Ruiz_framework} (further called Paper I), here we propose an alternative scaling approach based on \emph{similarity (or similitude) scaling}.  As discussed in Paper I, similarity scaling MagLIF loads preserves much of the physics regimes already known or being studied on today's Z pulsed-power driver.  By avoiding significant deviations into unexplored and/or less well-understood regimes, the risk of unexpected outcomes on future scaled-up experiments is reduced.  In this work, we shall derive the scaling rules for the experimental input parameters characterizing a MagLIF load, and we shall test the estimated scaling rules for various performance metrics against 2D {\color{black}rad-MHD} \textsc{hydra} simulations.\cite{Marinak:1996fs,Koning:2009}

\begin{figure}
	\begin{circuitikz}[american,scale=1] \draw
	(0,3) to[sinusoidal voltage source,l=$\varphi_{\rm oc}$,i=$I_s$] (0,0)
	(0,3) to[resistor,l=$Z_0$] (2,3)
  			to[inductor,l=$L_0$] (3.5,3) --(5,3)
  		    to[inductor,l=$L_1$] (8,3)
  		    to[vL,l_=$L_{\rm load}$,i=$I_l$] (8,0) -- (0,0) 
    (3.5,3) to[capacitor,l^=$C$,v=$\varphi_c$] (3.5,0)
    (5,0) to[vR,l_=$R_{\rm loss}$] (5,3);
	\end{circuitikz}
	\caption{{\color{black}Representative circuit diagram of the Z generator.}}
	\label{fig:scaling:circuit}
\end{figure}
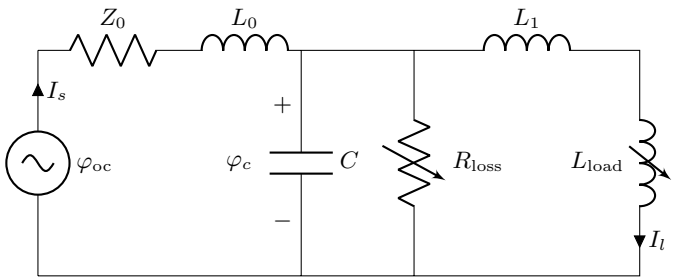

Scaling MagLIF loads to higher currents by using nondimensional-analysis was first proposed in \Refa{Schmit:2020jd}.  Both scaling approaches in Paper I and \Refa{Schmit:2020jd} leverage similarity scaling to derive scaling rules for the experimental input parameters of MagLIF loads.  However, the scaling rules derived in Paper I and \Refa{Schmit:2020jd} differ in two ways.  First, Paper I takes into account effects due to the changing liner thickness: the liner inner and outer radii do not follow the same scaling rules.  Therefore, the scaling laws for other fuel parameters, \eg, the density and the magnetic field, need to be modified to take this effect into account.  In consequence, the scaling rules for the fuel temperature and pressure are modified.  {\color{black}Second, Paper I proposes to conserve relative thermal-conduction losses modeled by an effective diffusion coefficient that follows a Bohm-like scaling on the electron Hall parameter.  The effective thermal-diffusion coefficient takes into account internal advection flows within the isobaric hot plasma as it comes into contact with the cold liner wall.  These flows may increase thermal and magnetic-flux losses, specially in the high-magnetization regime.\cite{Vekshtein:1983aa,Vekshtein:1986aa}}  This is in contrast to \Refa{Schmit:2020jd} {\color{black}where the Braginskii transport coefficient\cite{Braginskii:1965vl} for electron heat conduction was used to determine the scaling rule of the externally applied magnetic field.}  Overall, \Refa{Schmit:2020jd} is the first piece of work where the foundations of similarity current-scaling of MagLIF loads were laid down, and Paper I presents a refined scaling model based on that work.

This paper is organized as follows.  In \Sec{sec:scaling}, we derive the scaling rules of the input parameters for MagLIF when varying the peak current.  In \Sec{sec:numerical}, we introduce the numerical modeling tools used to test the similarity-scaling predictions and give the specific input parameters for the anchor load.  In \Sec{sec:implosion}, we compare the implosion dynamics of the similarity-scaled liners.  In \Sec{sec:stagnation}, we study the scaling rules for various metrics describing stagnation conditions.  In \Sec{sec:loss}, we discuss the scaling of the burn-width time of fusion yield and the energy-loss mechanisms.  In \Sec{sec:performance}, we test the theory predictions for various metrics measuring performance.  In \Sec{sec:conclusions}, we summarize our main results.  In \App{app:correction}, we discuss the origins of a correction factor introduced to the scaling rule of the liner outer radius.

\begin{figure}
 	\includegraphics[scale=.43]{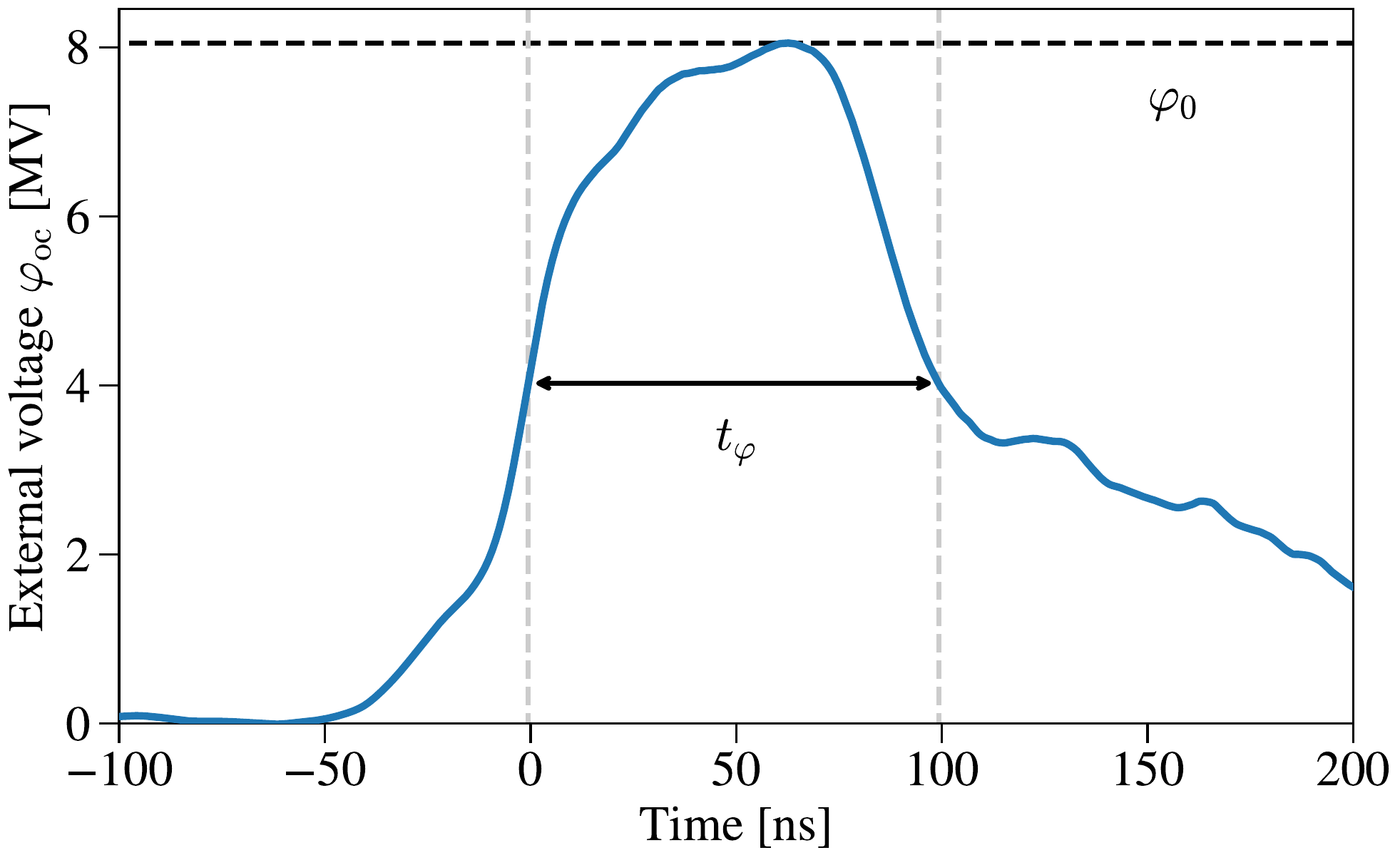}
	\caption{{\color{black}Example of a voltage source $\varphi_{\rm oc}$ as a function of time.  The characteristic voltage $\varphi_0$ is defined as the maximum value of $\varphi_{\rm oc}$.  The characteristic time $t_\varphi$ is defined as the full-width half-maximum (FWHM) of the voltage curve.}}
	\label{fig:scaling:Voc}
\end{figure}

%%%%%%%%%%%%%%%%%%%%%%%%%%%%%%%%%%%%%%%%%%%%%%%%%
%%%%%%%%%%%%%%%%%%%%%%%%%%%%%%%%%%%%%%%%%%%%%%%%%
%%%%%%%%%%%%%%%%%%%%%%%%%%%%%%%%%%%%%%%%%%%%%%%%%
\section{Current-scaling prescriptions}
\label{sec:scaling}

In Paper I, we derived the general framework for similarity-scaling MagLIF loads.\cite{foot:Ruiz_framework}  Here we shall focus on the scaling of MagLIF loads with respect to the characteristic current
\begin{equation}
	I_\star \doteq \frac{\varphi_0}{Z_0 +L_{\rm tot}/t_\varphi}.
	\label{eq:scaling:Istar}
\end{equation}
Here $\varphi_0$ is the characteristic voltage of the external voltage drive $\varphi_{\rm oc}(t)$ appearing in \Fig{fig:scaling:circuit} and can be defined as the maximum value of $\varphi_{\rm oc}(t)$.  The characteristic time $t_\varphi$ of the voltage drive can be defined as the full-width, half-maximum (FWHM) of the external voltage drive (see \Fig{fig:scaling:Voc}).  $Z_0$ is the impedance of the pulsed-power generator, and $L_{\rm tot}\doteq L_0+L_1$ is the total initial inductance of the circuit.  In this paper, we shall consider that the characteristic time $t_\varphi$ is constant.  Thus, all timescales are expected to be maintained--an assumption that will be tested throughout this paper.

%%%%%%%%%%%%%%%%%%%%%%%%%%%%%%%%%%%%%%%%%%%%%%%%%
\subsection{Scaling prescriptions for a MagLIF load}
\label{sec:scaling:MagLIF}

For the sake of completeness, we rewrite the scaling prescriptions of Paper I for the specific scaling scenario where only the characteristic current $I_\star$ is varied.  {\color{black}In Paper~I, two dimensionless parameters characterize the magnetic drive of the z-pinch implosion and the liner susceptibility towards instabilities:
\begin{align}
	\Pi  &	\doteq \frac{\mu_0 I_\star^2 }{4 \pi \widehat{m} R_{\rm out,0}^2 /  t_\varphi^2},
	\label{eq:scaling:Pi} \\
	\Psi  &	\doteq 2 \pi \frac{R_{\rm out,0}^2 \rho_{\rm ref}}{\widehat{m}} 
				\left( \frac{\mu_0 I_\star^2}{16 \pi^2 R_{\rm out,0}^2} \frac{1}{2 p_{\rm ref}} \right)^{2/\gamma}.
	\label{eq:scaling:Psi}
\end{align}
where $R_{\rm out,0}$ is the initial outer radius of the liner, $\widehat{m}$ is the liner mass per-unit-length, and $\mu_0$ is the magnetic permeability of free space.  $p_{\rm ref} $, $\rho_{\rm ref}$, and $\gamma$ are the reference pressure, mass density, and polytropic index that describe the equation-of-state (EOS) of an adiabatically compressed liner material, respectively.

The $\Pi$ parameter represents a ratio of the characteristic magnetic potential energy to the characteristic liner kinetic energy.  It denotes how strongly the magnetic drive accelerates the liner.\cite{Ryutov:2014hr}  The $\Psi$ parameter measures the susceptibility of the liner towards instabilities.\cite{Schmit:2020jd}  For thin-shell liners, the term $R_{\rm out,0}^2/\widehat{m}$ is proportional to the initial aspect ratio (AR$\doteq R_{\rm out,0}/\delta R_0$) of the liner.  Since the liner becomes more compressed and more unstable with higher magnetic pressures, $\Psi$ depends on the ratio $I_\star^2/R_{\rm out,0}^2$ which is proportional to the characteristic external magnetic pressure.  Hence, $\Psi$ increases with AR and magnetic pressure.  For large $\Psi$ values, the liner is more unstable.  (See Paper I for more details.)

We consider the parameters $p_{\rm ref} $, $\rho_{\rm ref}$, and $\gamma$ fixed.  When varying the characteristic current $I_\star$, we obtain two scaling laws for $R_{\rm out,0}$ and $\widehat{m}$:}
\begin{align}
	\frac{R_{\rm out,0}'}{R_{\rm out,0}}
			&=	
				\left( \frac{I_\star'}{I_\star} \right)^{\frac{\gamma-1}{2\gamma-1}}  
			 \simeq	
				\left( \frac{I_\star'}{I_\star} \right)^{\frac{\gamma-1}{2\gamma-1}} 
				\left[ 1 + \mc{C} \left( \frac{I_\star'}{I_\star} -1 \right) \right]	,
		\label{eq:scaling:Rout} \\
	\frac{\widehat{m}'}{\widehat{m}}
			 & =	\left( \frac{I_\star'}{I_\star} \right)^{2\gamma/(2\gamma-1)} .
		\label{eq:scaling:mhat} 
\end{align}
From hereon, for an arbitrary quantity $Q$ corresponding to the \emph{baseline} MagLIF load, the quantity $Q'$ denotes the value of the corresponding \textit{scaled} MagLIF load.  In \Eq{eq:scaling:Rout}, we included a correction term, which takes into account the differences in the liner shock-compression when changing the characteristic current driving the load.  (A further discussion on this topic is given in \App{app:correction}.)  Specific values for $\gamma$ and $\mc{C}$ are given in \Sec{sec:numerical}.  The scaling rules \eq{eq:scaling:Rout} and \eq{eq:scaling:mhat} guarantee that the liner will implode in a similar fashion and that its robustness towards instabilities will be maintained when increasing current.  After finding the scaled liner outer radius $R_{\rm out,0}'$ and its mass per-unit-length $\widehat{m}'$, we can determine the liner inner radius $R_{\rm in,0}'$ by using the definition of $\widehat{m}'$:
\begin{equation}
	R_{\rm in,0}' = \left[ R_{\rm out,0}'^2 - \widehat{m}'/(\pi \rho_{\rm liner,0}) \right]^{1/2},
	\label{eq:scaling:Rin}
\end{equation}
where $\rho_{\rm liner,0}$ is the initial density of the liner.  Upon knowing the liner dimensions of the baseline load and the parameters $\gamma$ and $\mc{C}$, \Eqs{eq:scaling:Rout}--\eq{eq:scaling:Rin} determine the scaling prescriptions for the radial dimensions of the liner.

As the liner implodes, the fuel pressure increases and eventually decelerates the liner.  In Paper I, this process is characterized by the dimensionless parameter $\Phi$,\cite{Schmit:2020jd} {\color{black}which is given by
\begin{equation}
	\Phi 	\doteq \frac{4}{3} \frac{E_{\rm preheat}  }{\widehat{m} h R_{\rm out,0}^2 /  t_\varphi^2},
			\label{eq:scaling:Phi}
\end{equation}
where $E_{\rm preheat}$ is the preheat energy delivered to the fuel and $h$ is the liner height.\cite{foot:linerheight}  The dimensionless parameter $\Phi$ measures the relative importance of the fuel preheat energy to the characteristic liner kinetic energy. When enforcing conservation of $\Phi$,} we find that the preheat energy per-unit-length $\smash{\widehat{E}_{\rm preheat}\doteq E_{\rm preheat}/h}$ scales as
\begin{equation}
	\frac{\widehat{E}_{\rm preheat}'}{\widehat{E}_{\rm preheat} }
		=	 \left( \frac{I_\star'}{I_\star} \right)^2.
	\label{eq:scaling:Epreheathat}
\end{equation}
In other words, the preheat energy per-unit-length $\smash{\widehat{E}_{\rm preheat}}$ scales as the square of the characteristic current $I_\star$ of the system.  Note that the \textit{total} preheat energy delivered to the fuel will scale as
\begin{equation}
	\frac{E_{\rm preheat}'}{E_{\rm preheat}}
		=	\frac{\widehat{E}_{\rm preheat}'}{\widehat{E}_{\rm preheat}}
			\frac{h'}{h}.
	\label{eq:scaling:Epreheat}
\end{equation}
Finally, since $t_\varphi = \const$ in this study, all timescales are assumed to be conserved.  Therefore, the time $t_{\rm preheat}$ at which preheat occurs remains unchanged:
\begin{equation}
		t_{\rm preheat}' = t_{\rm preheat} .
	\label{eq:scaling:tpreheat}
\end{equation}

Following Paper I, we scale the initial fuel density $\rho_0$, the external magnetic field $B_{z,0}$, and the liner height $h$ in order to conserve the relative radiation, {\color{black}thermal-conduction}, and end-flow energy losses, respectively.  {\color{black}Based on the  similarity-scaling framework, the dimensionless parameters characterizing these energy-loss mechanisms have the following dependencies on the MagLIF input parameters:
\begin{align}
	\Upsilon_{\rm rad} & \propto \frac{\rho_0^2 T_{\rm preheat}^{1/2} R_{\rm in,0}^2 h }{E_{\rm preheat}} \, t_\varphi, 	
		\label{eq:scaling:Upsilonrad} \\
	\Upsilon_c  & \propto \frac{\rho_0 T_{\rm preheat}^2 h }{E_{\rm preheat} B_{z,0} } \, t_\varphi, 	
		\label{eq:scaling:Upsilonc} \\
	\Upsilon_{\rm end} & \propto \frac{T_{\rm preheat}^{1/2} }{h}\, t_\varphi, 
		\label{eq:scaling:Upsilonend}
\end{align}
where $\smash{T_{\rm preheat}\propto E_{\rm preheat}/(\rho_0 R_{\rm in,0}^2 h)}$ is the characteristic fuel temperature achieved during preheat.  The dimensionless parameter \eq{eq:scaling:Upsilonc} considers thermal-conduction losses in a Bohm-like regime.  This is a result of internal advection flows arising in the isobaric, hot plasma core as the plasma is cooled by the cold liner walls.\cite{Vekshtein:1983aa,Vekshtein:1986aa}  Such advection flows enhance thermal-conduction losses, specially in the high-magnetization regime.\cite{Velikovich:2015gs}}  In the specific case of $t_\varphi=\const$, the scaling prescriptions for these quantities are given by
\begin{align}
	\frac{\rho_0'}{\rho_0}&
			=	\left( \frac{I_\star'}{I_\star} \right)^{2/3}
				\left( \frac{R_{\rm in,0}'}{R_{\rm in,0} } \right)^{-2/3}  ,
		\label{eq:scaling:rho}  \\
	\frac{B_{z,0}'}{B_{z,0}} &
			=	\left( \frac{I_\star'}{I_\star} \right)^{4/3}
				\left( \frac{R_{\rm in,0}'}{R_{\rm in,0} } \right)^{-10/3},
		\label{eq:scaling:Bz} \\
	\frac{h'}{h}&
			=	\left( \frac{I_\star'}{I_\star} \right)^{2/3}
				\left( \frac{R_{\rm in,0}'}{R_{\rm in,0} } \right)^{-2/3}.
		\label{eq:scaling:h} 
\end{align}
Note that the initial fuel density $\rho_0$ and the liner height $h$ follow the same scaling prescriptions.

{\color{black}Equations \eq{eq:scaling:Rout}--\eq{eq:scaling:Rin}, \eq{eq:scaling:Epreheathat}--\eq{eq:scaling:tpreheat}, and \eq{eq:scaling:rho}--\eq{eq:scaling:h} represent the scaling rules for the most important input parameters characterizing a MagLIF load.}  However, there are other specific features of a MagLIF load that are absent from the model introduced in Paper I.  As an example, other parameters defining the platform are the laser-spot size $R_{\rm spot}$ and the inner radius of the cushions $R_{\rm cushion}$.  [The cushions are cylindrical washers placed within the liner ends (see \Fig{fig:liners}) that help mitigate the wall instability,\cite{McBride:2013gda} which occurs where the liner meets the electrode surfaces.]  In Paper I, we did not invoke any models to describe the propagation of the preheat-induced blast-wave,\cite{HarveyThompson:2019ff} laser-plasma interactions,\cite{Geissel:2018ee} or the wall instability.\cite{McBride:2013gda}  For simplicity, here we shall invoke geometric similarity when scaling $R_{\rm spot}$ and $R_{\rm cushion}$.  In other words, we assume that these quantities scale proportionally to the initial inner radius of the liner:
\begin{equation}
	\frac{R_{\rm spot}'}{R_{\rm spot}} 
		= \frac{R_{\rm cushion}'}{R_{\rm cushion}} 
		=	\frac{R_{\rm in,0}'}{R_{\rm in,0}}.
	\label{eq:scaling:Rcushion}
\end{equation}
Likewise, other parameters such as the axial length of the cushions, the anode--cathode gap length, and the axial location of the laser-entrance-hole (LEH) window are geometrically scaled according to the liner height $h$.\cite{foot:axial_scaling}  We denote these axial dimensions by $H$, and the scaling rule is then
\begin{equation}
	\frac{H'}{H} = \frac{h'}{h}.
	\label{eq:scaling:H}
\end{equation}
These additional scaling rules complete the scaling prescriptions for the input parameters defining MagLIF.

{\color{black}
%%%%%%%%%%%%%%%%%%%%%%%%%%%%%%%%%%%%%%%%%%%%%%%%%
\subsection{Scaling prescriptions for the circuit parameters}
\label{sec:scaling:circuit}

The circuit model for the pulsed-power generator shown in \Fig{fig:scaling:circuit} is used to drive the simulated MagLIF implosions.  In \Fig{fig:scaling:circuit}, $\varphi_{\rm oc}(t)$ is the external time-varying drive voltage and is twice the forward-going voltage at the vacuum-insulator stack on Z.  (An example time-trace for $\varphi_{\rm oc}$ is shown in \Fig{fig:scaling:Voc}.)  $Z_0$ is the effective impedance of the pulsed-power generator, $L_0$ is the inductance of the outer magnetically-insulated transmission lines, $\varphi_c(t)$ is the corresponding voltage across the capacitor $C$ associated to the MITLs, $L_1$ is the initial inductance of post-convolute feed region, $R_{\rm loss}(t)$ is a shunt resistor using a prescribed time-dependent model, and $L_{\rm load}(t)$ is the time-varying inductance of the imploding MagLIF load.  The model used for the shunt resistor is given by
\begin{equation}
	R_{\rm loss} (t)  \doteq R_{\rm loss,i} + (R_{\rm loss,f}- R_{\rm loss,i}) f(t),
	\label{eq:electric:Rloss}
\end{equation}
where
\begin{equation}
	f(t) = \frac{1}{1+ \exp\left( - \frac{t-t_{\rm loss}}{\Delta t_{\rm loss}} \right) } .
	\label{eq:electrical:f}
\end{equation}
is a function describing the transition from the initial loss resistance $R_{\rm loss,i}$ of the circuit early in time to the final loss resistance $R_{\rm loss,f}$ at later times. Specific values for the circuit components are given in \Sec{sec:numerical}.

Section~II of Paper I provides the governing equations for the circuit dynamics.  When rewriting the equations in dimensionless form, six dimensionless parameters appear describing the circuit inductance matching, the LR-circuit drive efficiency, the LC-circuit resonance, relative current losses, and the load--circuit coupling.  These parameters are given by
\begin{equation}
	\begin{aligned}
	c_1 &\doteq \frac{L_0}{L_0+L_1},  							& c_2 & \doteq \frac{Z_0 t_\varphi}{L_0+L_1},  \\
	c_3 &\doteq (L_0+L_1)^{1/2} \frac{C^{1/2}}{t_\varphi}, 	& c_4 &	\doteq \frac{R_{\rm loss,i} t_\varphi}{L_0+L_1},   \\
	c_5 &\doteq \frac{R_{\rm loss,f} t_\varphi}{L_0+L_1}, 		& c_6 & \doteq \frac{\mu_0 h}{2\pi(L_0+L_1)}.
	\end{aligned}
	\label{eq:electrical:variables}
\end{equation}

Since the characteristic timescale $t_\varphi$ is fixed, the circuit parameters scale according to the load height $h$:}
\begin{equation}
	\frac{Z_0'}{Z_0}  
		=	\frac{L_0'}{L_0} 
		=	\frac{L_1'}{L_1} 
		=	\frac{R_{\rm loss,i}'}{R_{\rm loss,i}} 
		=	\frac{R_{\rm loss,f}'}{R_{\rm loss,f}} 
		= 	\frac{h'}{h},
	\label{eq:scaling:Z} 
\end{equation}
\begin{equation}
	\frac{C'}{C} 
		 = \frac{h}{h'}.
	\label{eq:scaling:C} 
\end{equation}
As shown, all inductances and resistances appearing in the electrical circuit scale proportionally to the liner height $h$.  Only the capacitance $C$ scales inversely proportionally to $h$.  Since $t_\varphi$ remains constant, all timescales appearing in the problem are maintained.  Hence, 
\begin{equation}
	t_{\rm loss}' = t_{\rm loss},
	\qquad
	\Delta t_{\rm loss}' = \Delta t_{\rm loss}.
	\label{eq:scaling:tloss} 
\end{equation}

{\color{black}When considering the characteristic current $I_\star$ as the independent scaling variable and substituting the scaling relations in \Eqs{eq:scaling:Z} into \Eq{eq:scaling:Istar}, we find that} the external voltage drive scales proportionally to $I_\star$ and the liner height $h$:
\begin{equation}
	\frac{\varphi_0'}{\varphi_0}
		=	\frac{I_\star'}{I_\star} \frac{L_0'}{L_0}
		=	\frac{I_\star'}{I_\star} \frac{h'}{h}.
	\label{eq:scaling:varphi} 
\end{equation}
This scaling rule is only valid when the characteristic time $t_\varphi$ is held constant.  In order to increase the peak current by a fraction $I_\star'/I_\star$, the voltage drive will have to be multiplied by the factor $\varphi_0'/\varphi_0$ given in \Eq{eq:scaling:varphi}.  From \Eq{eq:scaling:h}, we know that MagLIF liners change in axial length when scaling the current.  Hence, $\varphi_0$ does not scale linearly with the characteristic current $I_\star$ (or peak current, for that matter).  Instead, it shows a stronger scaling, which translates to higher voltage requirements for scaled-up MagLIF loads.

As a final remark of this section, when scaling MagLIF loads and the circuit parameters according to the presented scaling prescriptions, it is expected that the normalized current delivered to the load $\bar{I}_l \doteq I_l/I_{\star}$ will remain invariant. (This will be tested in \Sec{sec:implosion}.)  Therefore, the peak current delivered to the load $\I \doteq \max(I_l)$ scales linearly with $I_\star$; in other words, $\I'/\I = I_\star'/I_\star$.  Since peak current $\I$ is a commonly used metric for current delivery to z-pinch devices, we shall express the scaling laws in terms of $\I$ in the rest of this paper.

%%%%%%%%%%%%%%%%%%%%%%%%%%%%%%%%%%%%%%%%%%%%%%%%%
%%%%%%%%%%%%%%%%%%%%%%%%%%%%%%%%%%%%%%%%%%%%%%%%%
%%%%%%%%%%%%%%%%%%%%%%%%%%%%%%%%%%%%%%%%%%%%%%%%%
\section{Numerical simulations and baseline load parameters}
\label{sec:numerical}

We conducted 2D \textsc{hydra} simulations to test the similarity-scaling theory.  \textsc{hydra} is a massively parallel arbitrary Lagrangian--Eulerian (ALE) radiation, resistive-diffusion, magneto-hydrodynamics code\cite{Marinak:1996fs,Koning:2009} and is one of the main design tools for MagLIF experiments.\cite{Sefkow:2014ik,HarveyThompson:2018dd,Weis:2021id}  For the calculations presented in this paper, the simulations were performed in cylindrical geometry with azimuthal symmetry.  A generalized Ohm's law was used that includes effects such as Nernst advection, which can affect the magnetization of the fuel.  The equation of state and the transport coefficients for the nonideal thermal and magnetic conduction of the DT fuel, Be liner, Al electrodes, and stainless-steel cushions were taken from pregenerated LEOS and SESAME tables.\cite{More:1988jx,foot:SESAME} The radiation field was modeled using implicit Monte-Carlo photonics.

{\color{black}The 2D calculations presented in this paper are ``clean". In other words, they do not include impurity mixing into the DT fuel nor random initial seeding of the magneto-Rayleigh--Taylor (MRT) instability\cite{Harris:1962hu,Weis:2015hk,Velikovich:2015jl} on the outer surface of the liner.  As a result, the simulation predictions of performance metrics, such as fusion yield, are inherently optimistic.  Nevertheless, our goal is to utilize the 2D calculations to test the scaling rules derived from the similarity-scaling theory.}

\begin{figure}
	\includegraphics[scale=0.44]{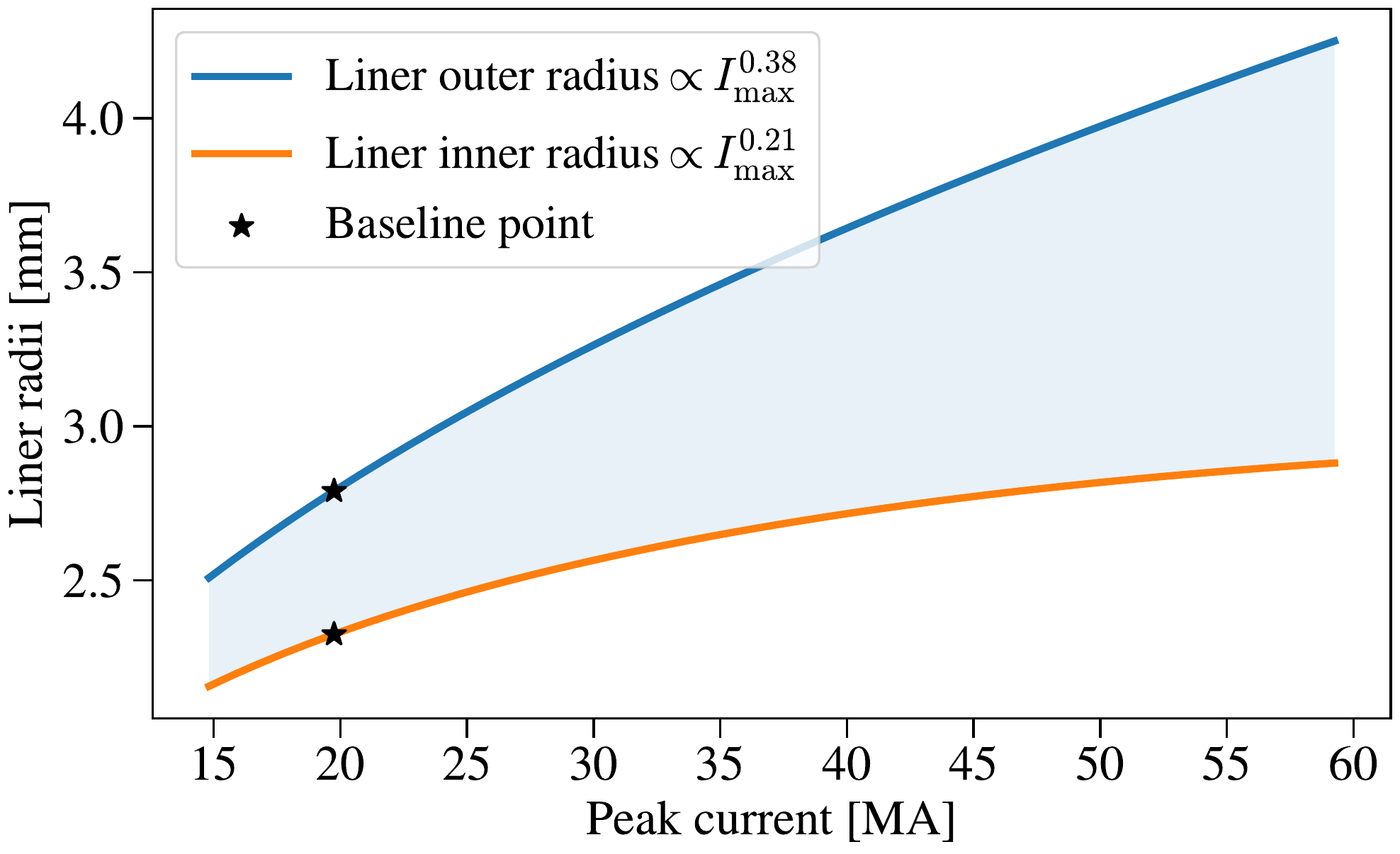}
	\caption{Scaling curves for the initial inner and outer radii of a MagLIF liner.  These curves are based on a typically fielded MagLIF liner with $R_{\rm out,0} =2.79$~mm and AR=6 driven at 20-MA peak current.\cite{Gomez:2020cd}  The curves are obtained from \Eqs{eq:scaling:Rout}--\eq{eq:scaling:Rin}.  Shaded region denotes the area where the MagLIF liner is initially located.}
	\label{fig:numerical:radii}
\end{figure}

\begin{figure*}
	\includegraphics[scale=0.43]{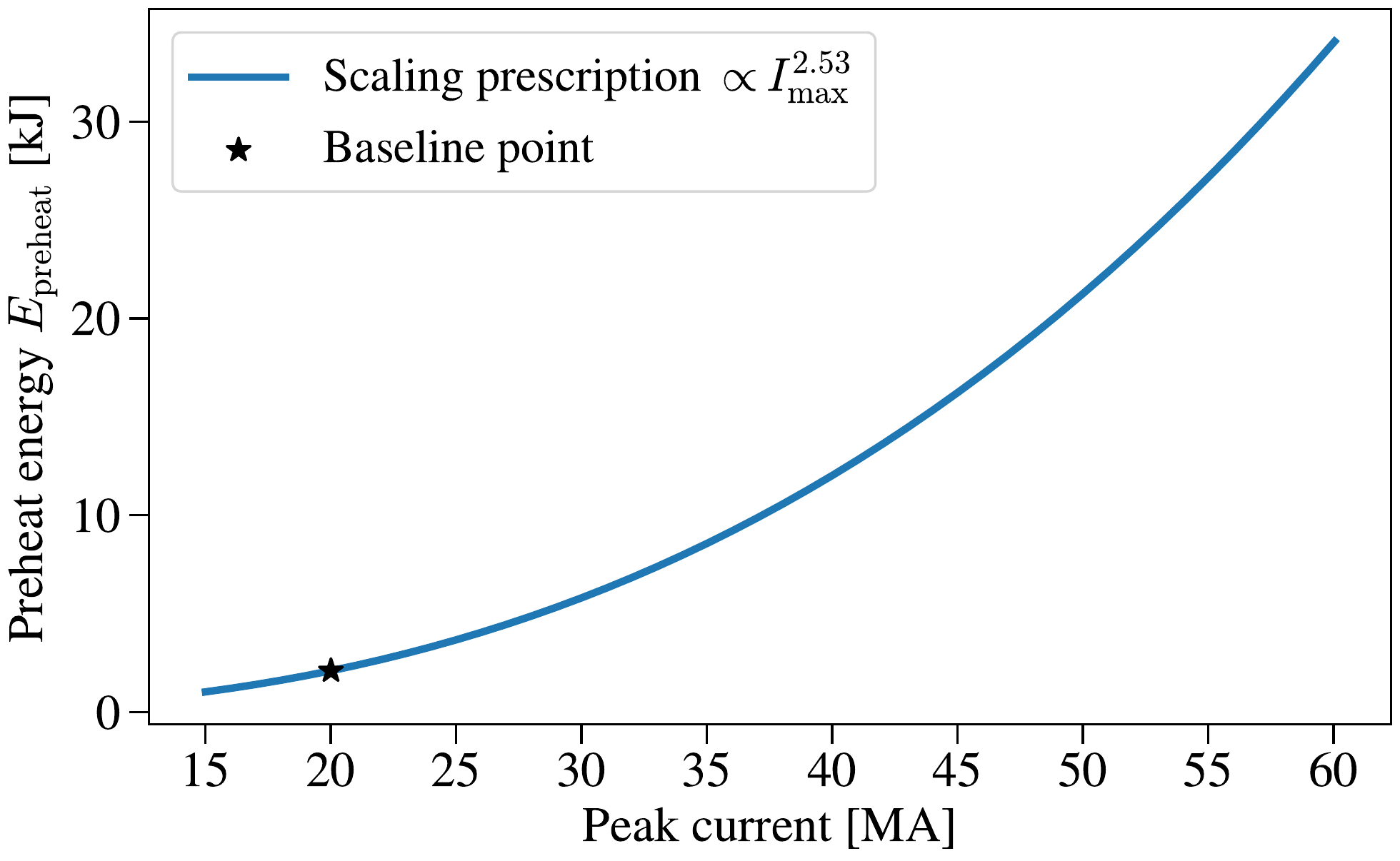}
	\hspace{0.3cm}
	\includegraphics[scale=0.43]{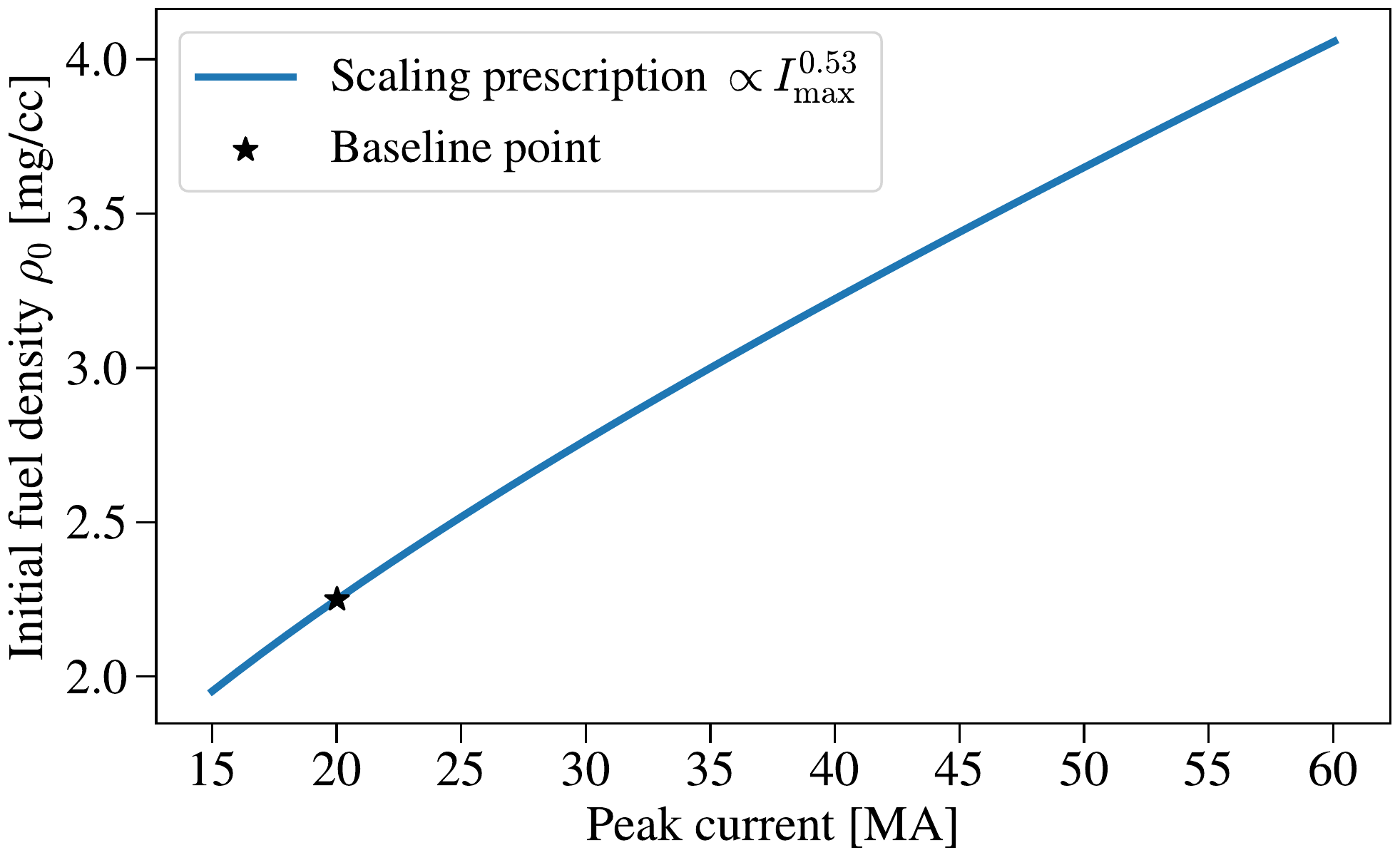}
	\includegraphics[scale=0.43]{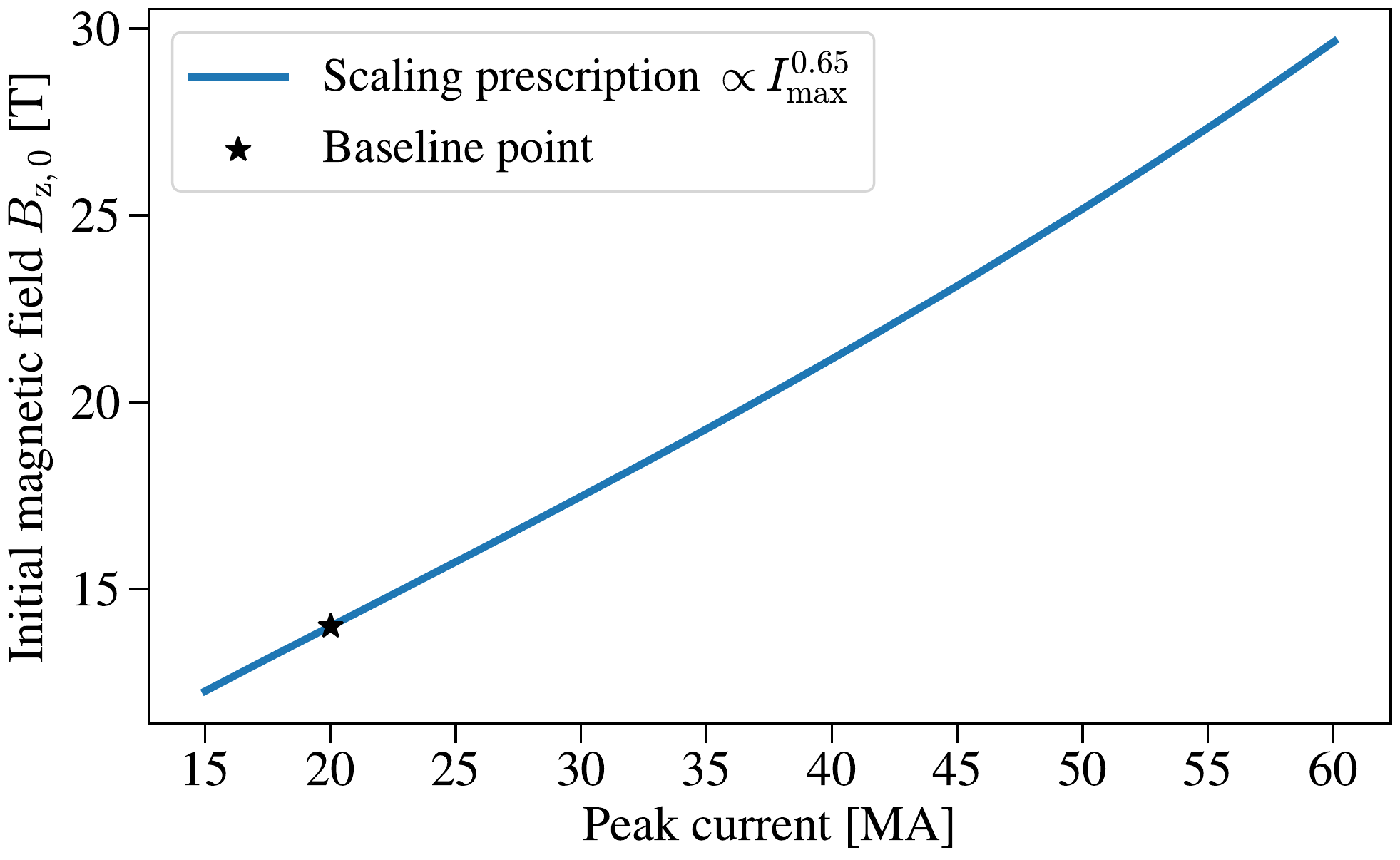}
	\hspace{0.3cm}
	\includegraphics[scale=0.43]{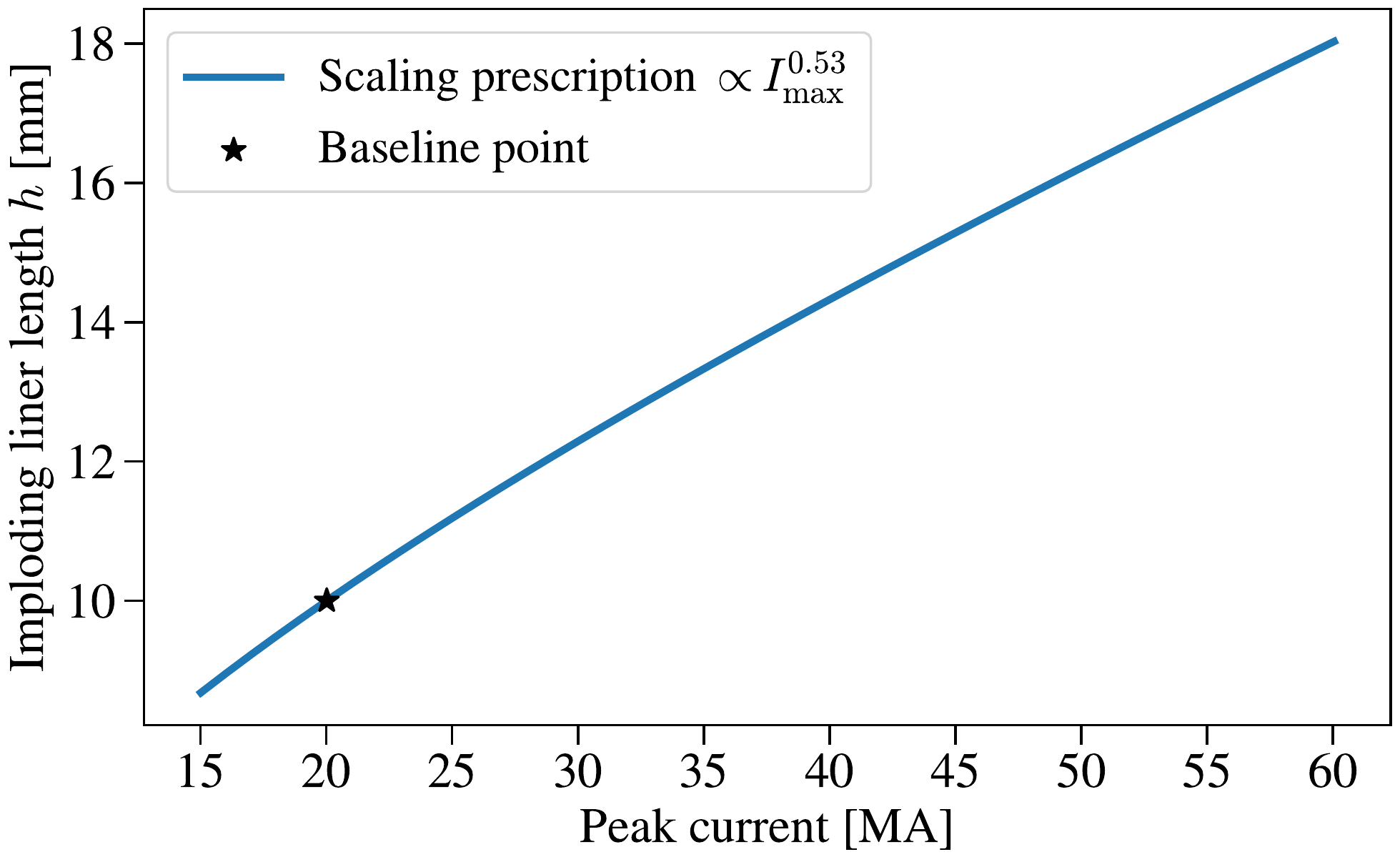}
	\caption{Scaling curves for the preheat energy, initial fuel density, initial magnetic field, and liner height.  The legends in the subfigures give the approximate power-law fits to the scaling prescriptions.}
	\label{fig:numerical:parameters}
\end{figure*}

The simulations are externally driven using the circuit model shown in \Fig{fig:scaling:circuit}.  The parameters for the baseline circuit model are $Z_0=0.18~\Omega$, $L_0 = 9.58$~nH, $C = 0.1$~nF, and $L_1 = 5$~nH.  For the shunt resistor, we use $R_{\rm loss,i} = 80$~Ohm and  $R_{\rm loss,f} = 0.25$~Ohm.\cite{foot:circuit}  The time parameters for the shunt resistor are $t_{\rm loss} = 0$~ns and  $\Delta t_{\rm loss} = 5$~ns.  The 20-MA, baseline load is driven using the open-source voltage shown in \Fig{fig:scaling:Voc}.  The circuit parameters and the voltage drive were scaled according to \Eqs{eq:scaling:Z}--\eq{eq:scaling:varphi}.  

For the baseline MagLIF configuration, we consider an initial liner outer radius of $R_{\rm out,0}=2.79$~mm and an initial inner radius of $R_{\rm in,0}=2.325$~mm.  Thus, the initial aspect ratio AR$\doteq R_{\rm out,0}/(R_{\rm out,0}-R_{\rm in,0})$ of the liner is six.   In simulations, the liner is made of Be with initial density 1.858~g/cm$^3$, so the mass per-unit-length is approximately 139~mg/cm.  Regarding the initial fuel parameters, we consider an equimolar DT gas fill at $\rho_0 = 2.25$~mg/cm$^3$ density.  As a reminder, the Z facility does not presently have the capability of fielding MagLIF loads with equimolar DT fuel.  When maintaining the initial number of electrons in the fuel, $\rho_0$ corresponds to 1.8~mg/cm$^3$ of DD fuel.  The preimposed initial axial magnetic field is {\color{black}$B_{z,0}=14$~T}.  For the preheat energy deposition, the fuel is heated uniformly by adding 2.1~kJ of energy into a plasma column of radius $R_{\rm spot} = 0.75$~mm coaxial to the liner.  The deposition of energy begins approximately 70~ns before burn time and lasts for about 10 ns.\cite{foot:preheat}  The liner height is 10 mm.  With the exception of the slightly higher density and preheat energy, these chosen parameters are representative of the input parameters of MagLIF loads typically fielded in present-day experiments on Z.\cite{Gomez:2020cd,YagerElorriaga:2022cp}

In this paper, we consider a polytropic index for the Be liner of $\gamma = 2.25$ in \Eqs{eq:scaling:Rout} and \eq{eq:scaling:mhat}.  This value is slightly larger than a fit to the cold EOS curve $\gamma_{\rm cold}  \simeq 1.9$.  We use this higher value for $\gamma$ in order to account for the shock heating of the liner and the ensuing larger incompressibility of the liner shell.  {\color{black} It is important to note that $\gamma$ serves as a semi-empirical parameter that represents an ``effective" polytropic index characterizing the liner compressibility for the ensemble of similarity-scaled implosions studied in this paper.  As noted in Section III of Paper I, self-consistently obtaining an adiabatic polytropic index $\gamma$ based on tabular EOS for Be is difficult since the trajectories in the EOS phase-space are highly dependent on the liner-implosion dynamics, and in most cases, MagLIF liners are shocked by the magnetic pressure drive. Future work may include improving the present scaling study by replacing the simple adiabatic EOS model with constant $\gamma$ with a more sophisticated and complete material EOS model.}

The correction factor $\mc{C}$ in \Eq{eq:scaling:Rout} was chosen to be {\color{black}$\mc{C} = 0.02/(48/20-1)\simeq1.4\%$}.  For a MagLIF load driven at a peak current close to 48~MA, this correction factor denotes that the scaled outer radius of the liner is shifted outwards by 1.4$\%$ compared to its nominal scaled value without the correction.  {\color{black}Including this correction term allows to better conserve the implosion trajectories of the scaled MagLIF loads.  A discussion of the liner-implosion dynamics is presented in \Sec{sec:implosion}.}

With the scaling prescriptions in \Eqs{eq:scaling:Rout}--\eq{eq:scaling:Rin} and the parameters given in the preceding paragraphs, we plot in \Fig{fig:numerical:radii} the initial inner and outer radii of the similarity-scaled MagLIF loads.  The scaling law for $R_{\rm in,0}$ in \Eq{eq:scaling:Rin} follows a complex dependency on $R_{\rm in,0}$ and $\widehat{m}$.  {\color{black}To simplify the upcoming analysis, we fitted a power-law to the liner radii within the range shown in \Fig{fig:numerical:radii}. The resulting approximate power-law scaling rules for the liner radial dimensions are the following:
\begin{equation}
	\frac{R_{\rm out,0}'}{R_{\rm out,0}} \simeq \left( \frac{\I'}{\I} \right)^{0.381} , \qquad
	\frac{R_{\rm in,0}'}{R_{\rm in,0}} \simeq \left( \frac{\I'}{\I} \right)^{0.206}.  
	\label{eq:numerical:R}
\end{equation}
}
As shown from the equations above, when increasing the characteristic current (or equivalently, the peak current), the liner becomes larger in radius.  This is mainly a consequence of constraining the liner to implode in a similar fashion by conserving the $\Pi$ {\color{black}parameter in \Eq{eq:scaling:Pi}.}  Since $R_{\rm in,0}$ grows more slowly compared to $R_{\rm out,0}$, the liner becomes significantly thicker when scaling to higher currents.  To be more quantitative, the initial aspect ratio AR for the anchor liner equals 6 while the AR for the corresponding scaled 60-MA liner is close to 3.1.  The increase of the liner thickness is a consequence of {\color{black}conserving the $\Psi$ parameter in \Eq{eq:scaling:Psi}} to maintain the robustness of the liner towards the MRT instability.\cite{Harris:1962hu,Weis:2015hk,Velikovich:2015jl,Sinars:2010de,McBride:2012db,McBride:2013gda,Awe:2014gba,Ruiz:2022aa}

The scaling rules for the preheat energy $E_{\rm preheat}$, the initial fuel density $\rho_0$, the applied axial magnetic field $B_{z,0}$, and the liner height $h$ are shown in \Fig{fig:numerical:parameters}.  {\color{black}As in \Eqs{eq:numerical:R}, when fitting the exact scaling prescriptions \eq{eq:scaling:Epreheat} and \eq{eq:scaling:rho}--\eq{eq:scaling:h} to power laws, we find the following scaling relations:}
\begin{gather}
	\frac{E_{\rm preheat}'}{E_{\rm preheat}} \simeq \left( \frac{\I'}{\I} \right)^{2.529} , 
		\label{eq:numerical:Epreheathat}\\
	\frac{\rho_0'}{\rho_0} = \frac{h'}{h}  \simeq \left( \frac{\I'}{\I} \right)^{0.529}, 
		\label{eq:numerical:rho}\\
	\frac{B_{z,0}'}{B_{z,0}}   \simeq \left( \frac{\I'}{\I} \right)^{0.647} .
		\label{eq:numerical:Bz}
\end{gather}
Therefore, a preheat energy of 2.1 kJ at 20-MA peak current scales to 34 kJ at 60 MA.  From \Eqs{eq:scaling:rho} and \eq{eq:scaling:h}, the scaling rules for the initial fuel density and the liner height are identical.  When scaling between 20~MA and 60~MA, we find that the initial fuel density $\rho_0$ increases from 2.25 mg/cm$^3$ to 4.1 mg/cm$^3$.  To mitigate end losses, the liner height also increases substantially from 10~mm to 18.3~mm.  To maintain relative thermal ion-conduction losses, the externally preimposed magnetic field must increase from 14~T to 30~T.  

\begin{figure}
	\includegraphics[scale=.45]{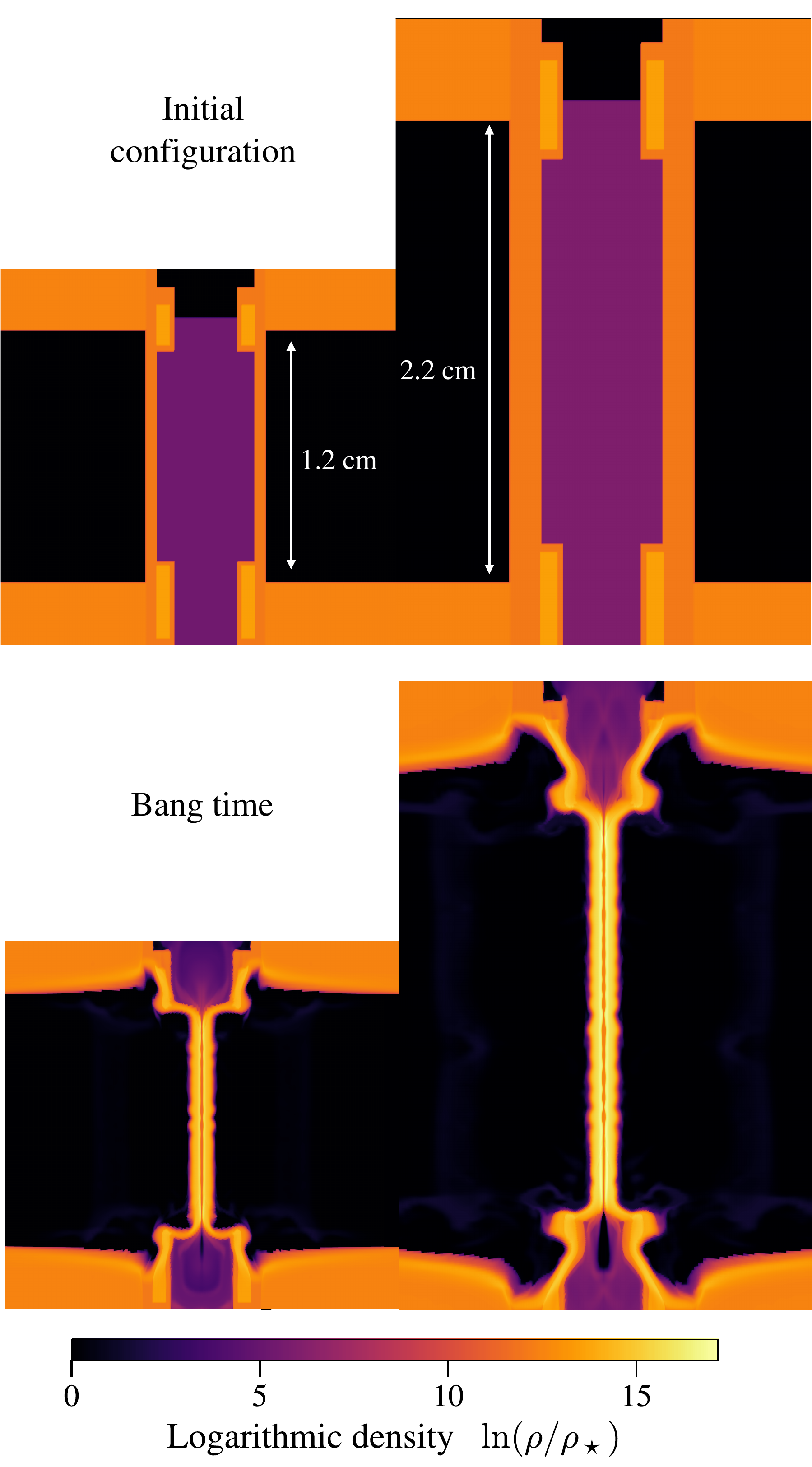}
	\caption{Top left: Logarithmic density plot for the anchor AR=6 MagLIF load driven at a 20-MA peak current.  Top right:  Similarity scaled, AR=3.1 MagLIF load driven at 60-MA peak current.  Bottom:  Corresponding logarithmic density plots near stagnation calculated using \textsc{hydra}.  {\color{black}Density maps are normalized to $\rho_\star = 10^{-5}$~g/cc.}}
	\label{fig:liners}
\end{figure}

Figure~\ref{fig:liners} (top) presents logarithmic-density plots of the initial configurations of the baseline 20-MA MagLIF load and of a similarity-scaled 60-MA load.  When increasing the current drive, MagLIF liners become larger in radius, taller, and thicker.  Figure~\ref{fig:liners} (bottom) illustrates that the similarity-scaled MagLIF loads look qualitatively the same near stagnation.  This is a signature of similarity scaling.  In Secs.~\ref{sec:implosion}--\ref{sec:performance}, we shall present a quantitative comparison of the implosion dynamics, the stagnation conditions of the plasma fuel, and the performance of the scaled MagLIF loads.

At this point, it is worth commenting on the differences between the power-law scaling rules in \Eqs{eq:numerical:R}--\eq{eq:numerical:Bz} and the scaling rules proposed in \Refa{Schmit:2020jd}.  In this paper, we focus on current scaling and keep the characteristic time of the voltage source constant.  Therefore, the scaling rules discussed here correspond to the ``implosion-time conserving, radiation-conserving" (ITC-rad) scaling strategy of \Refa{Schmit:2020jd}.  When neglecting the small correction added to take into account shock-compression effects, the scaling rules for the liner radial dimensions are identical.  {\color{black}However, in this work, we take into account the finite-thickness of the liner when deriving the scaling rules for the fuel-related quantities.  This amounts to considering $R_{\rm in,0}$ in the scaling laws instead of $R_{\rm out,0}$ as in \Refa{Schmit:2020jd}.  In consequence,} the scaling prescriptions for the initial gas density and the liner height increase slightly more rapidly (\eg, $\rho_0 \propto \I^{0.53}$ in this paper versus $\rho_0 \propto \I^{0.42}$ in \Refa{Schmit:2020jd}).  In addition, \Refa{Schmit:2020jd} suggests to keep $B_{z,0}$ constant when increasing peak current since electron-conduction losses decrease.  {\color{black}However, Paper I invokes a thermal-loss model that follows a more conservative Bohm-like scaling.}  This leads to the $B_{z,0}\propto \I^{0.65}$ scaling prescription in \Eq{eq:numerical:Bz}.  {\color{black}Because considering $R_{\rm in,0}$ gives a more accurate scaling law for the fuel volume, which then increases more slowly with current,} the scaling rules in this paper lead to hotter and higher-pressure stagnation columns, which then modify the scaling laws for important performance metrics, for example, the fusion yield.

\begin{figure}
	\includegraphics[scale=.43]{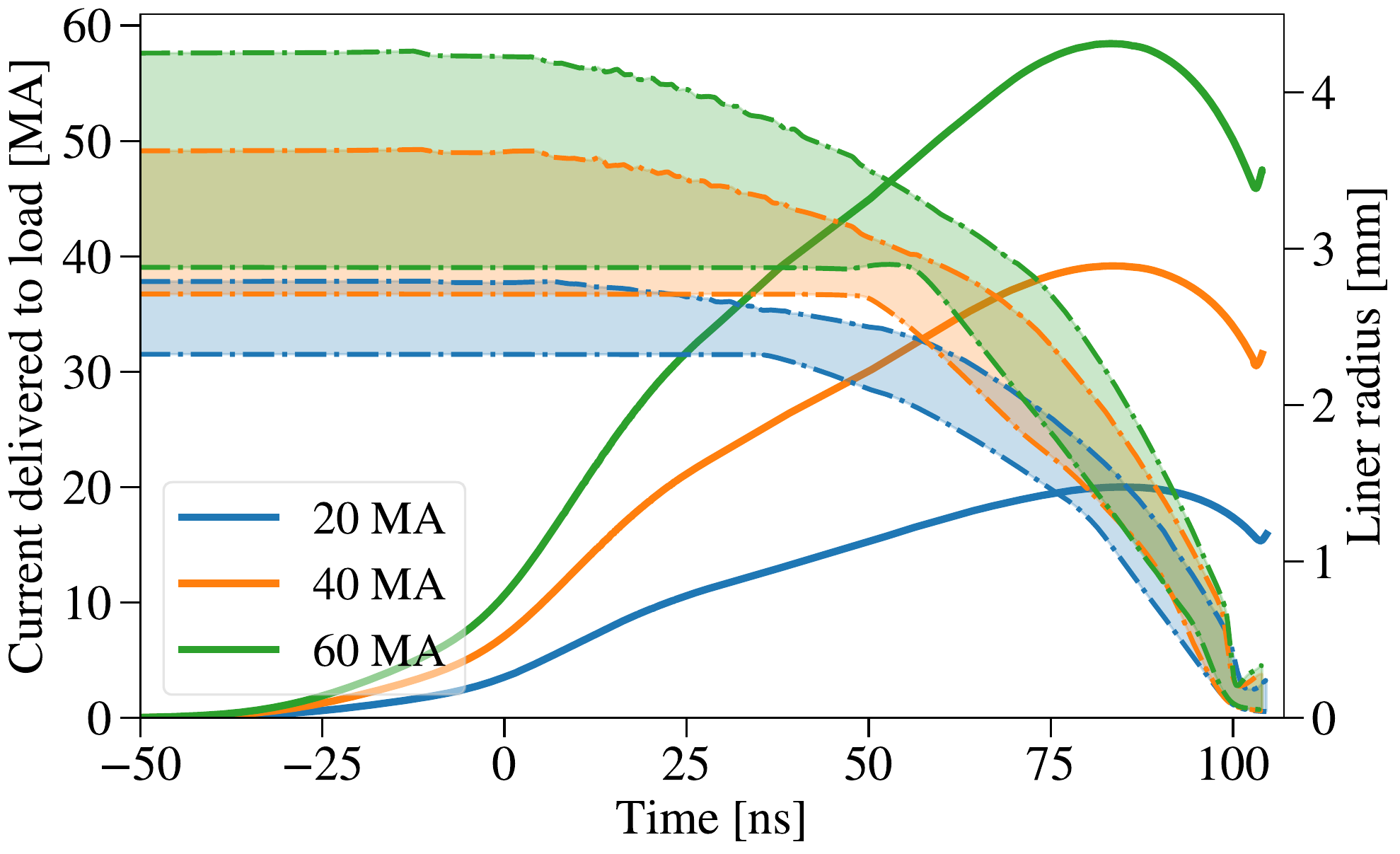}
	\caption{Liner radii and current delivered to the load versus time for different current scales calculated using \textsc{hydra}.  The 20-MA anchor load (shown in blue) stagnates at the same time as its similarity-scaled counterparts driven at 40~MA and 60~MA (shown in orange and green, respectively).  The outer boundary of the liner is tracked using a $1/e\simeq37\%$ threshold of the maximum density (similar to \Refa{Bose:2017jf}), and the inner boundary is tracked using a Lagrangian marker.  {\color{black}The liner radii curves shown here are the result of averaging over the liner axial length.}}
	\label{fig:radius_vs_time}
\end{figure}

%%%%%%%%%%%%%%%%%%%%%%%%%%%%%%%%%%%%%%%%%%%%%%%%%
%%%%%%%%%%%%%%%%%%%%%%%%%%%%%%%%%%%%%%%%%%%%%%%%%
%%%%%%%%%%%%%%%%%%%%%%%%%%%%%%%%%%%%%%%%%%%%%%%%%
\section{Liner-implosion dynamics}
\label{sec:implosion}

Figure \ref{fig:radius_vs_time} shows the radius-versus-time and current-versus-time plots for three similarity scaled MagLIF loads driven at 20~MA, 40~MA, and 60~MA.  As shown from the shaded regions in \Fig{fig:radius_vs_time}, the liners of the scaled MagLIF loads tend to be larger in radius and thicker.  When adopting the scaling prescriptions in \Sec{sec:scaling} for the liner radial dimensions and the electrical circuit, we find that all three loads implode at similar times according to {\color{black}2D clean} \textsc{hydra} calculations.  This is not surprising because all timescales are expected to be conserved.

\begin{figure}
	\includegraphics[scale=.43]{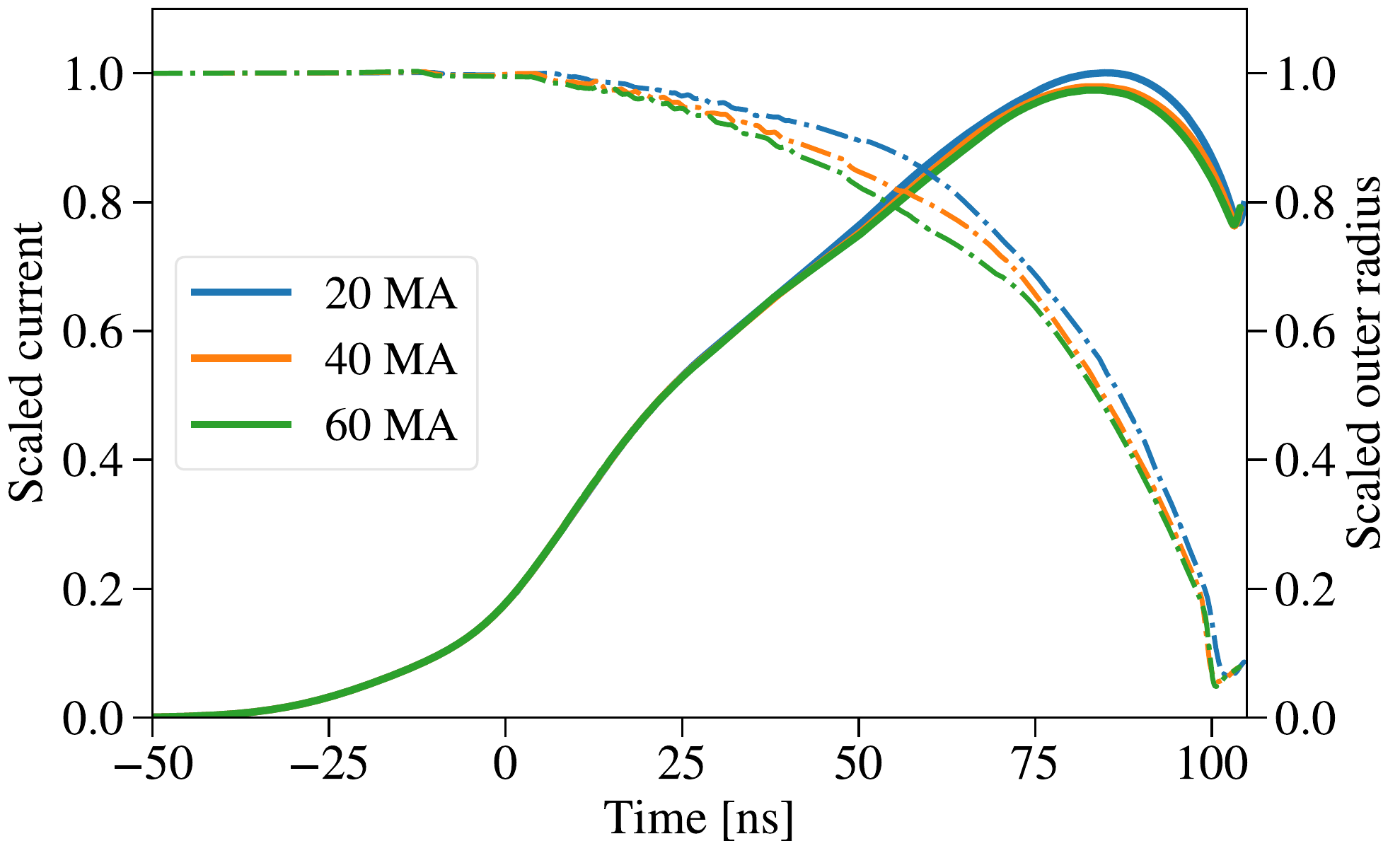}
	\caption{Normalized implosion trajectory of the outer liner radius and normalized current traces calculated using \textsc{hydra}.  When following the scaling prescriptions provided in \Sec{sec:scaling}, the normalized trajectories of the outer liner radius are almost identical.  Likewise, the normalized current traces are also in very close agreement, as expected from the scaling theory.}
	\label{fig:radius_norm_vs_time}
\end{figure}

In \Fig{fig:radius_norm_vs_time}, we normalize the liner outer radius by its initial value so that $\bar{R}_{\rm out}(t)\doteq R_{\rm out}(t)/R_{\rm out}(0)$ is plotted.  The currents delivered to the loads are also normalized by 20, 40,~and~60~MA, which are the expected scaled peak currents.  {\color{black}As shown in \Fig{fig:radius_norm_vs_time}, the trajectories of $\bar{R}_{\rm out}$ remain invariant due to similarity.}  The normalized current delivered to the load $\bar{I}_l$ is almost perfectly scale invariant indicating that the scaling prescriptions in \Sec{sec:scaling} for the electrical circuit and the liner radial dimensions hold.

\begin{figure}
	\includegraphics[scale=.43]{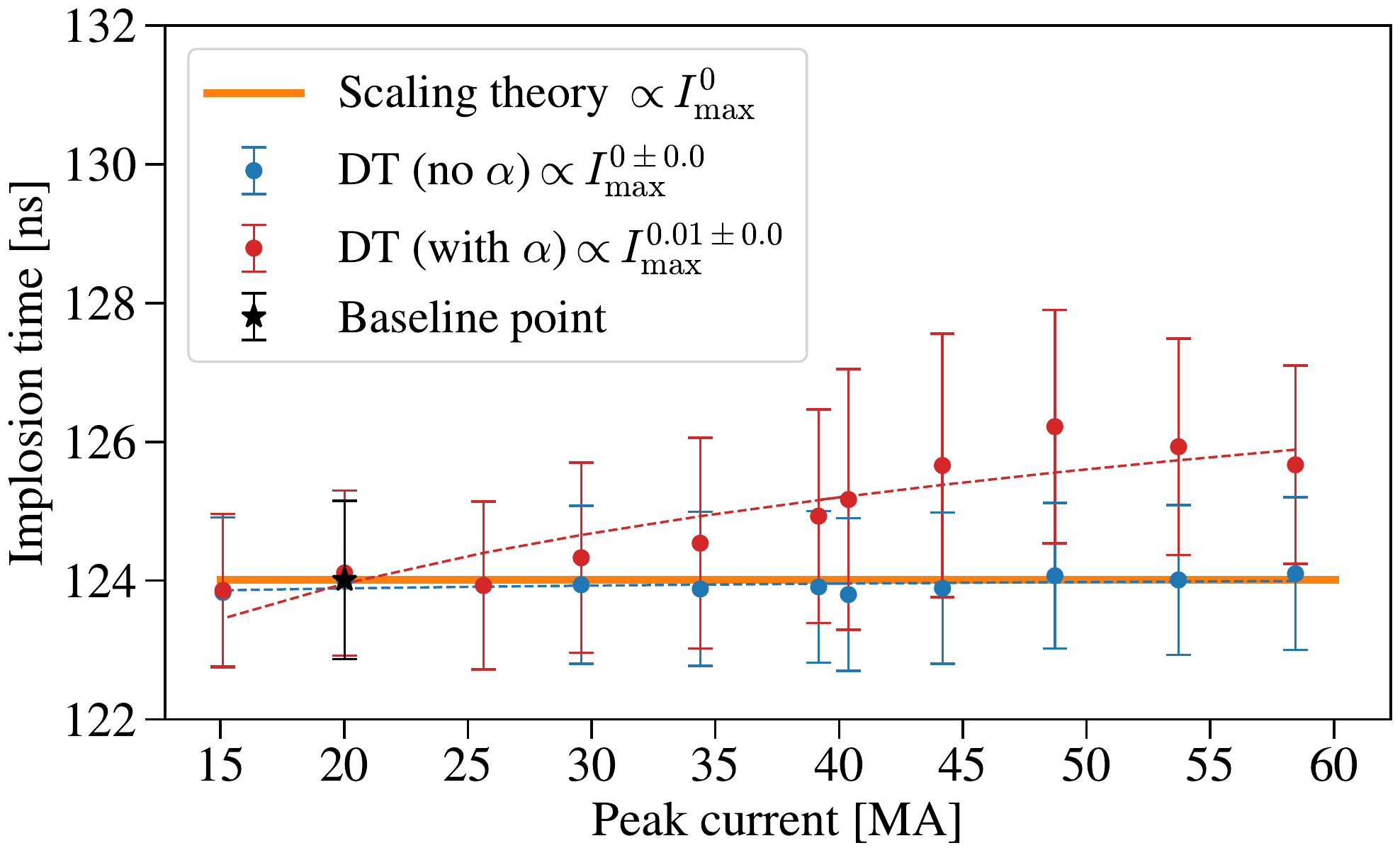}
	\caption{Implosion times of the similarity-scaled MagLIF loads.  Red and blue points denote simulation results with and without $\alpha$ heating, respectively.  Both datasets have the same experimental input parameters.  For reference purposes, the error bars denote the full-width, half-maximum burn time of the neutron-production event in the simulations.
}
	\label{fig:implosion_time}
\end{figure}

To provide a more quantitative comparison of similarity between liner implosions, we plot in \Fig{fig:implosion_time} the simulated implosion times\cite{foot:timplosion} for a family of MagLIF loads {\color{black}scaled from 15 MA to 60 MA.}  For the simulations without $\alpha$ heating, the implosion time is conserved.  {\color{black}The good agreement between the implosion times is due in part by the correction factor $\mc{C}\simeq1.4\%$ included \Eq{eq:scaling:Rout}.  Without the correction factor $(\mc{C}=0)$, the 60-MA configuration implodes 2 ns earlier than the baseline 20-MA configuration.  This deviation is small and of the order of one burn-width time.  When including the correction, the discrepancy in the implosion time between the two calculations reduces to a fraction of a nanosecond.  The correction $\mc{C}$ allows to adjust the implosion times for the scaled-up configurations and only slightly changes the scaling prescriptions for the MagLIF input parameters.} Finally, concerning the calculations with $\alpha$ heating, the deviation in the implosion time is about one burn-width time.

The in-flight aspect ratio (IFAR) is often used as a measure of the robustness of ICF shell implosions towards Rayleigh--Taylor instabilities.\cite{Bose:2017jf}  Higher IFAR values are usually correlated to less stable implosions.  Figure \ref{fig:IFAR_vs_time} shows the IFAR trajectories plotted versus time for the MagLIF loads shown in \Fig{fig:radius_vs_time}.  The IFAR increases during the early stages of the implosions due to the shock compression of the liners.  After shock breakout, the liners then relax and accelerate as a whole.  This occurs roughly when the outer convergence ratio $\mathrm{CR}_{\rm out}(t) \doteq R_{\rm out,0}/R_{\rm out}(t)$ has reached a value of 1.5 or close to $\sim$75 ns in simulation time.  From \Fig{fig:IFAR_vs_time}, it is clear that the initial AR of the scaled-up liners must decrease in order to compensate for the stronger magnetic compression of the liner.  This design feature was not taken into account in previous scaling works.\cite{Slutz:2016cf,Slutz:2018iq}  Note that the peak IFAR values for the similarity-scaled liners are smaller than that of the 20-MA baseline load.  Therefore, the scaling prescriptions \eq{eq:numerical:R} for the liner dimensions obtained using $\gamma=2.25$ could be considered slightly ``over-conservative" with respect to robustness of the liner towards instabilities.  This is a favorable feature since performance of MagLIF implosions can be significantly degraded by MRT instabilities in simulations.

The scaling rules in \Sec{sec:scaling} do not guarantee that every physical process in a MagLIF implosion will be conserved.  One example of such a process is the strength of the blast wave occurring after preheat.  Figure~\ref{fig:radius_vs_time} shows that the inner radius of the 60-MA scaled liner slightly increases at around 55~ns, which is the time when the blast wave impacts the liner.  This effect is not visible for the loads with lower preheat energy.  The increase in the blast-wave strength may lead to unaccounted interface mixing between the fuel and the liner during the preheat stage.  Figure~\ref{fig:radius_norm_vs_time} also shows that the normalized outer radius of the scaled 60-MA load is more strongly magnetically compressed (near 50 ns) before the liner begins to accelerate rapidly.  This is understandable since the magnetic pressure driving the liners scales as 
\begin{equation}
	\frac{p_{\rm mag,ext}'}{p_{\rm mag,ext}}
		= \left( \frac{\I'}{\I} \frac{R_{\rm out,0}}{R_{\rm out,0}'} \right)^2
		\simeq \left( \frac{\I'}{\I} \right)^{1.24},
\end{equation}
so the higher-current liners are subject to stronger shock compression.  {\color{black}This increases the load inductance and reduces the load current, which explains the small reduction in the normalized current at 50~ns shown in \Fig{fig:radius_norm_vs_time} when comparing the 20-MA and 60-MA time traces.}  Overall, the similarity-scaling framework presented in Paper I will not conserve all the physics involved in a MagLIF implosion.  However, this framework can preserve the \emph{leading-order} physical processes and provide reasonable estimates of the scaled performance metrics.

%%%%%%%%%%%%%%%%%%%%%%%%%%%%%%%%%%%%%%%%%%%%%%%%%
%%%%%%%%%%%%%%%%%%%%%%%%%%%%%%%%%%%%%%%%%%%%%%%%%
%%%%%%%%%%%%%%%%%%%%%%%%%%%%%%%%%%%%%%%%%%%%%%%%%
\section{Stagnation conditions}
\label{sec:stagnation}

In this section, we examine the scaling rules for the fuel thermodynamic conditions that are achieved near peak burn and compare these against simulation results.  {\color{black}From Paper~I, we recall that any ``no-alpha" dimensionless dynamical quantity $\smash{\bar{Q}_{\rm no\,\alpha}(\bar{t})}$ can be approximately written as a function of the dimensionless parameters defining a MagLIF load:
\begin{equation}
	\bar{Q}_{\rm no\,\alpha} 
		\simeq \mathcal{F}^{(0)}_{\bar{\varphi}_{\rm oc},f} (\bar{t}; c_{1-6},\bar{t}_i, \Pi,\Phi,\Psi,\Upsilon_{\rm rad},\Upsilon_c,\Upsilon_{\rm end}),
	\label{eq:stagnation:general}
\end{equation}
where $\mathcal{F}^{(0)}_{\bar{\varphi}_{\rm oc},f}$ depends on the dimensionless time trace $\bar{\varphi}_{\rm oc}$ of the voltage drive and on the function $f$ in \Eq{eq:electrical:f} parameterizing the shunt resistor $R_{\rm loss}$.  $\bar{t}_i$ denotes the dimensionless time parameters, \eg, $\bar{t}_{\rm loss}$, $\Delta \bar{t}_{\rm loss}$ and $\bar{t}_{\rm preheat}$.  (For further details, see the discussion provided in Sec.~X~A of Paper~I.)  The dimensionless parameters appearing on the right-hand side of \Eq{eq:stagnation:general} are conserved when adopting the scaling prescriptions in \Sec{sec:scaling}.  In consequence, the right-hand side remains invariant across current scales \textit{for similarity-scaled MagLIF configurations}.  Hence,
\begin{equation}
	\bar{Q}_{\rm no\,\alpha}'(\bar{t}) \simeq \bar{Q}_{\rm no\,\alpha}(\bar{t}),
	\label{eq:stagnation:qbar}
\end{equation}
where $\bar{Q}$ and $\bar{Q}'$ denote the dimensionless quantities corresponding to a \textit{baseline} and a \textit{scaled} MagLIF configurations, respectively.  Equation~\eq{eq:stagnation:qbar} has already been demonstrated for the particular case of the dimensionless liner implosion trajectories shown in \Fig{fig:radius_norm_vs_time}.

\begin{figure}
	\includegraphics[scale=.43]{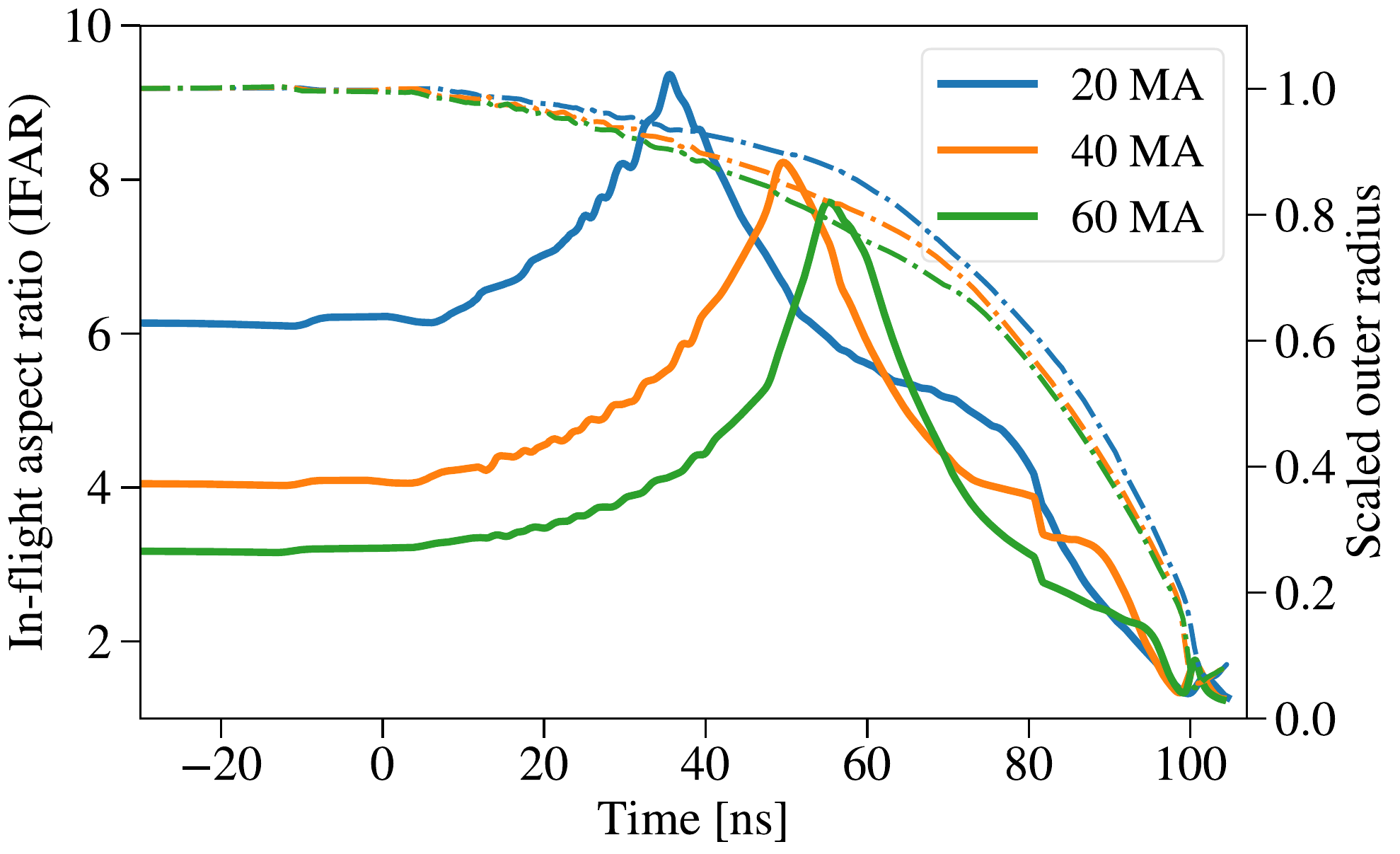}
	\caption{IFAR trajectories plotted versus time for the MagLIF loads shown in \Fig{fig:radius_vs_time}.  Before the inner surface of the liner has moved, the initial shock transiting the liner decreases its thickness and increases the IFAR.  Once shock breakout occurs, the liner relaxes and begins to move as a whole.}
	\label{fig:IFAR_vs_time}
\end{figure}

In Paper I, the dimensionless variables $\bar{Q}$ are constructed by normalizing the dimensional quantities $Q$ by functions of the known experimental input parameters; \ie $\bar{Q}=Q/Q_0$ where $Q_0$ is a normalization coefficient that only depends on input parameters, \eg, $R_{\rm in,0}$, $\rho_0$, and $E_{\rm preheat}$.  As an example, the fuel pressure $p_{\rm fuel}$ is normalized by the preheat pressure $p_{\rm preheat}\doteq (2/3) E_{\rm preheat} /(\pi R_{\rm in,0}^2 h)$, which is the characteristic fuel pressure achieved by the preheat.  From \Eq{eq:stagnation:qbar}, the scaling rules for the dimensional variables $Q$ are given by
\begin{equation}
	\frac{Q_{\rm no\,\alpha}'}{Q_{\rm no\,\alpha}} \simeq \frac{Q_0'}{Q_0}.
	\label{eq:stagnation:q}
\end{equation}
Once a baseline quantity $Q_{\rm no\,\alpha}$ is known (calculated via simulations or measured in experiments), the corresponding scaled quantity $Q'$ is obtained by multiplying $Q$ by a known function $Q_0'/Q_0$ that depends on the two sets of input parameters of the similarity-scaled MagLIF configurations.  Equations~\eq{eq:stagnation:qbar} and \eq{eq:stagnation:q} are only valid for similarity-scaled MagLIF configurations.  In the following, we shall make use of \Eqs{eq:stagnation:qbar} and \eq{eq:stagnation:q} to derive the scaling rules for the fuel thermodynamic conditions near stagnation, and we shall compare the scaling rules against 2D clean \textsc{hydra} simulation results.}

\begin{figure}
	\includegraphics[scale=.43]{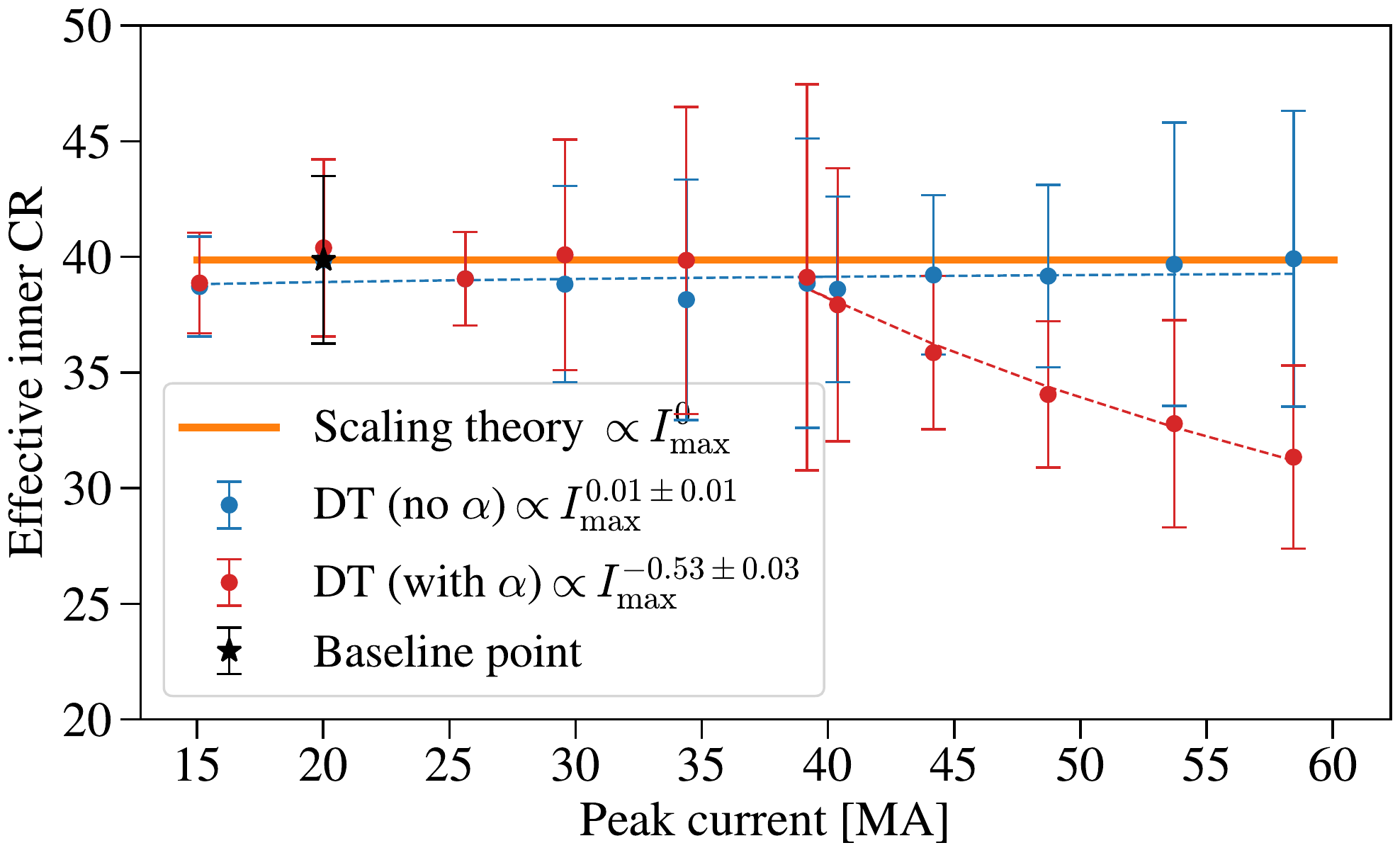}
	\caption{Effective inner convergence ratio $\mathrm{CR}_{\rm in,eff}$ evaluated as defined in \Eq{eq:CReff} for the similarity-scaled MagLIF loads.  Red and blue points denote simulation results with and without $\alpha$ heating, respectively.  Both data sets have the same experimental input parameters.  Error bars correspond to a standard deviation when collecting statistics of the radial position of the {\color{black}tracked fuel--liner interface along the axial length.}}
	\label{fig:CR}
\end{figure}

In the absence of energy-loss mechanisms, the inner-convergence ratio $\mathrm{CR}_{\rm in}(t) \doteq R_{\rm in,0} / R_{\rm in}(t)$ is the main factor determining the thermodynamic conditions of the fuel plasma.  When scaling MagLIF loads to higher currents, it is desirable that $\mathrm{CR}_{\rm in}$ near peak burn be maintained since higher convergence ratios often correlate with more unstable plasma columns at stagnation.  Since peak burn may occur before or after peak compression of the fuel column (depending on the relative importance of $\alpha$ heating), we introduce the effective inner convergence ratio:
\begin{equation}
	\mathrm{CR}_{\rm in,eff} 
	\doteq  \left\{
             \begin{array}{ll}
               \mathrm{CR}_{\rm in}(t_{\rm bang}),\qquad t_{\rm bang} \leq t_{\rm stag} \\                
               \mathrm{CR}_{\rm in}(t_{\rm stag}),\qquad ~t_{\rm bang} > t_{\rm stag}
             \end{array}
              \right. ,
    \label{eq:CReff}
\end{equation}
where $t_{\rm bang}\doteq \argmax(\dot{Y})$ is the time at which peak burn occurs, $\smash{\dot{Y}(t) \doteq \mathrm{d}Y/\mathrm{d}t}$ is the neutron yield rate, and $t_{\rm stag} \doteq \argmax({\rm CR}_{\rm in})$ is the time at which peak compression of the fuel occurs.  This measure of the inner convergence ratio may be more representative of the risks associated to hydrodynamical instabilities affecting the burn event.  {\color{black}Since $\mathrm{CR}_{\rm in,eff}$ is a dimensionless dynamical quantity, we expect that $\mathrm{CR}_{\rm in,eff}$ should be conserved for the scaled loads without $\alpha$ heating included; \ie
\begin{equation}
	\mathrm{CR}_{\rm in,eff}' \simeq \mathrm{CR}_{\rm in,eff}.
\end{equation}
}
As shown in \Fig{fig:CR}, $\mathrm{CR}_{\rm in,eff}$ for the no-$\alpha$ calculations is maintained within error bars, thus indicating that the scaled-up MagLIF loads are not converging more.  Interestingly, simulations with $\alpha$ heating show a reduction of $\mathrm{CR}_{\rm in,eff}$ when going beyond 40-MA peak current.  As we shall discuss later on, calculations of similarity-scaled MagLIF loads suggest that $\alpha$ heating becomes more important for peak currents greater than 40~MA.  The heat source from the $\alpha$ particles lead to higher fuel pressures causing the fuel to stagnate at lower $\mathrm{CR}_{\rm in,eff}$ values.

\begin{figure}
	\includegraphics[scale=.43]{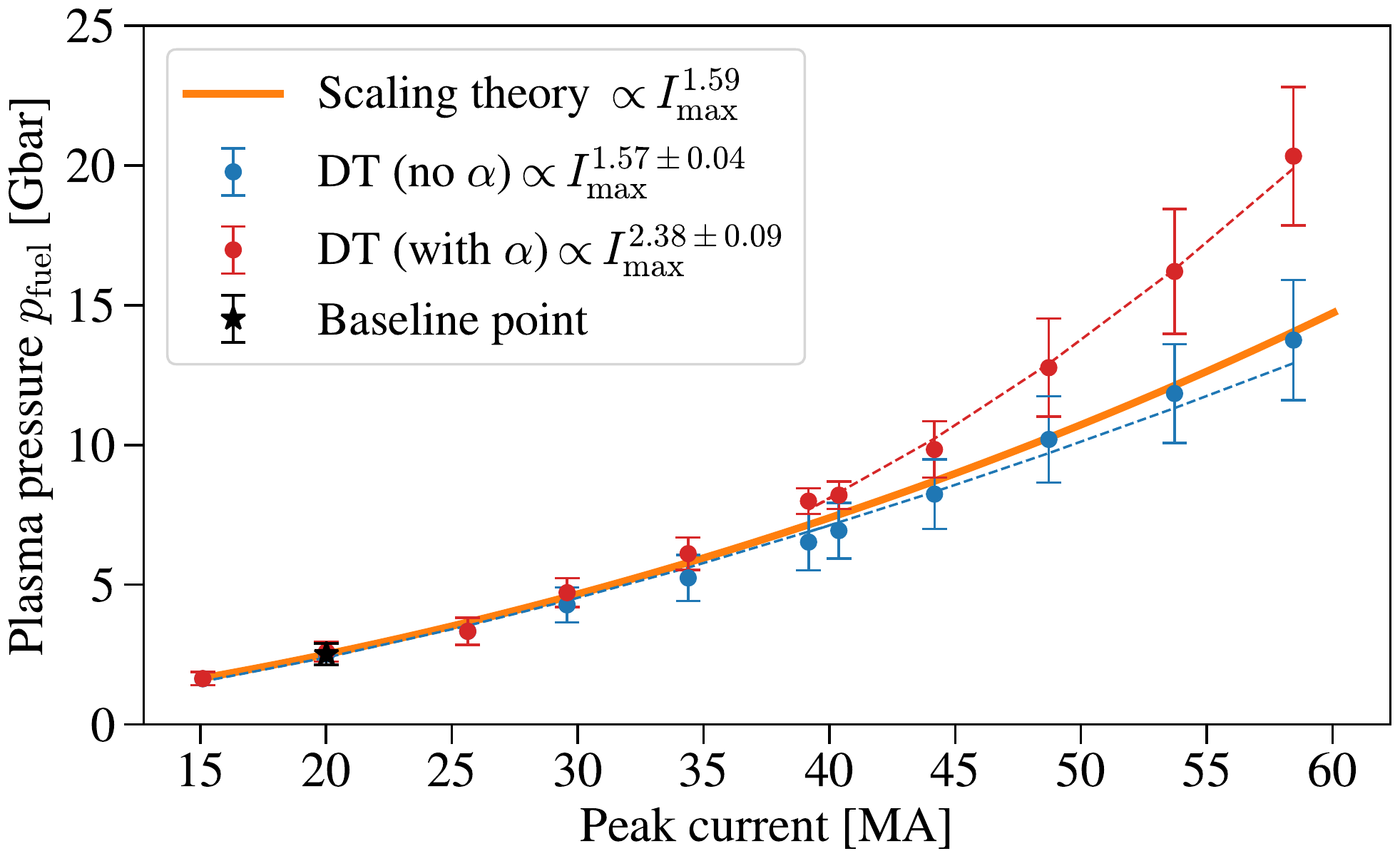}
	\caption{Burn-history averaged plasma pressure.  Red and blue points denote simulation results with and without $\alpha$ heating, respectively.  Dashed lines are power-law fits to the simulation data.  The legend shows the fitted scaling exponents.  Error bars denote the burn-weighted standard deviation associated to temporal variations of the plasma pressure near peak burn.  The orange curve is the theoretical scaling law in \Eq{eq:stagnation:pion2}.  }
	\label{fig:pion}
\end{figure}

Figure \ref{fig:pion} shows the plasma pressure averaged over the burn history.  In this paper, quantities $Q$ averaged over the burn history are calculated as follows:
\begin{equation}
	\langle Q \rangle_{\rm b.h.} 
		\doteq 
			\frac{ \int \, \int_{V_{\rm fuel}} n_i^2 \langle \sigma v \rangle \, Q \, 
						\mathrm{d}V \, \mathrm{d}t }
		                {\int \int_{V_{\rm fuel}} n_i^2 \langle \sigma v \rangle \, \mathrm{d}V \, \mathrm{d}t},
	\label{eq:stagnation:average}
\end{equation}
where $n_i^2 \langle \sigma v \rangle$ is proportional to the neutron yield rate per-unit-volume and $V_{\rm fuel}(t)$ is the volume of the fuel plasma.  In Paper I, the fuel pressure is normalized by the preheat pressure $p_{\rm preheat} \doteq (2/3)\smash{E_{\rm preheat} /(\pi R_{\rm in,0}^2 h) }$.  Thus, the no-$\alpha$ fuel pressure satisfies the scaling relation:
\begin{equation}
	\frac{p_{\rm fuel,no \, \alpha}'}{p_{\rm fuel,no \, \alpha}} 
			\simeq \frac{p_{\rm preheat}'}{p_{\rm preheat}} 
			= \frac{\widehat{E}_{\rm preheat}'}{\widehat{E}_{\rm preheat}} 
				\left( \frac{R_{\rm in,0}}{R_{\rm in,0}'}\right)^2  .
	\label{eq:stagnation:pion}
\end{equation}
Upon using the derived scaling rules in \Eqs{eq:scaling:Epreheathat} and \eq{eq:numerical:R}, we find that the plasma pressure approximately scales as
\begin{equation}
	\frac{p_{\rm fuel,no \, \alpha}'}{p_{\rm fuel,no \, \alpha}} 
		 \simeq \left( \frac{\I'}{\I}\right)^{1.59}.
	\label{eq:stagnation:pion2}
\end{equation}
%2-0.206*2=1.59
Equation~\eq{eq:stagnation:pion2} shows good agreement with the burn-history averaged, no-$\alpha$ fuel pressures of the similarity-scaled MagLIF loads shown in \Fig{fig:pion}.  {\color{black}Interestingly, when increasing the peak current from 20 MA to 60 MA, the no-$\alpha$ pressure is expected to increase by a factor of $(60/20)^{1.59}\simeq 5.7$.} Since $\alpha$ heating is not a process that is conserved, the simulation results with $\alpha$ heating show a stronger scaling law for the fuel pressure.  When the drive current exceeds 40-MA peak current, the power law fit for the simulation outputs is $p_{\rm fuel,\alpha} \propto \I^{2.33}$.  This stronger scaling curve is a signature of {\color{black}$\alpha$ heating effects becoming more prominent.}

\begin{figure}
	\includegraphics[scale=.43]{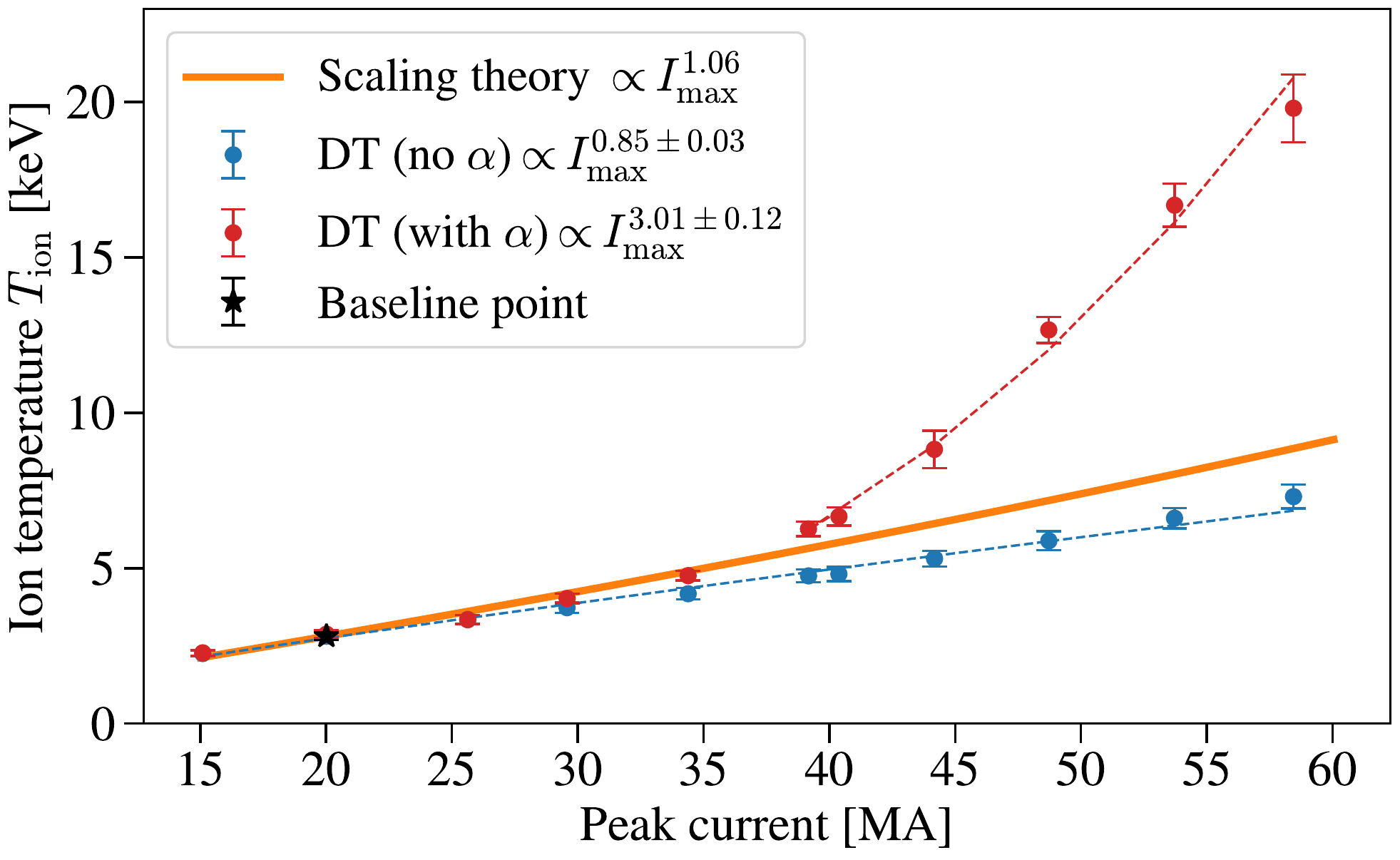}
	\caption{Burn-history averaged ion temperature.  Dashed lines are power-law fits to the simulation data.  Error bars denote the burn-weighted standard deviation associated with temporal variations of the ion temperature near peak burn.  The orange curve is the theoretical scaling curve \eq{eq:stagnation:tion2}.}
	\label{fig:tion}
\end{figure}

To further constrain the plasma thermodynamic conditions near stagnation, we compare the burn-history averaged fuel temperature $\langle T \rangle_{\rm b.h.}$.  Similar to the fuel pressure, the no-$\alpha$ fuel temperature scales as the preheat temperature $k_B T_{\rm preheat} \doteq p_{\rm preheat}/(2 \rho_0 / m_i)$.  We obtain
\begin{equation}
	\frac{T_{\rm no \,\alpha}'}{T_{\rm no \,\alpha}} 
			\simeq \frac{T_{\rm preheat}'}{T_{\rm preheat}} 
			= 	\frac{p_{\rm preheat}'}{p_{\rm preheat}} 
				\frac{\rho_0}{\rho_0'} 
			= \left(  \frac{\I'}{\I}\frac{R_{\rm in,0}}{R_{\rm in,0}'}\right)^2
				\frac{\rho_0}{\rho_0'}  ,
	\label{eq:stagnation:tion}
\end{equation}
where we used \Eq{eq:stagnation:pion}.  Upon substituting the scaling prescriptions of \Sec{sec:numerical}, we obtain the approximate power-law scaling rule for the no-$\alpha$ fuel temperature:
\begin{equation}
	\frac{T_{\rm no \,\alpha}'}{T_{\rm no \,\alpha}} 
		 \simeq \left(  \frac{\I'}{\I} \right)^{1.06}.
	\label{eq:stagnation:tion2}
\end{equation}
%%2-0.206*2 -0.529=1.06
The no-$\alpha$ fuel temperature is expected to grow approximately linearly with peak current when following the scaling rules proposed in this paper.

Figure~\ref{fig:tion} compares the theoretical scaling law to the simulation results.  In this case, the scaling theory slightly overpredicts the growth of the no-$\alpha$ ion temperatures.  (This discrepancy will be discussed in \Sec{sec:loss}.)  For the 20-MA baseline configuration, $\langle T_{\rm ion} \rangle_{\rm b.h.} = 2.9$~keV and increases to $\langle T_{\rm ion} \rangle_{\rm b.h.} =7.6$~keV at 60 MA, which exceeds the temperature threshold needed to have $\alpha$ heating dominate radiation losses.  As shown in the same figure, once the peak current exceeds 40 MA, calculations with $\alpha$ heating show that the fuel temperature markedly increases with a fitted scaling curve of $T_{\rm ion,\alpha} \propto \I^{2.8}$ and can reach 21~keV at 60~MA.

\begin{figure}
	\includegraphics[scale=.43]{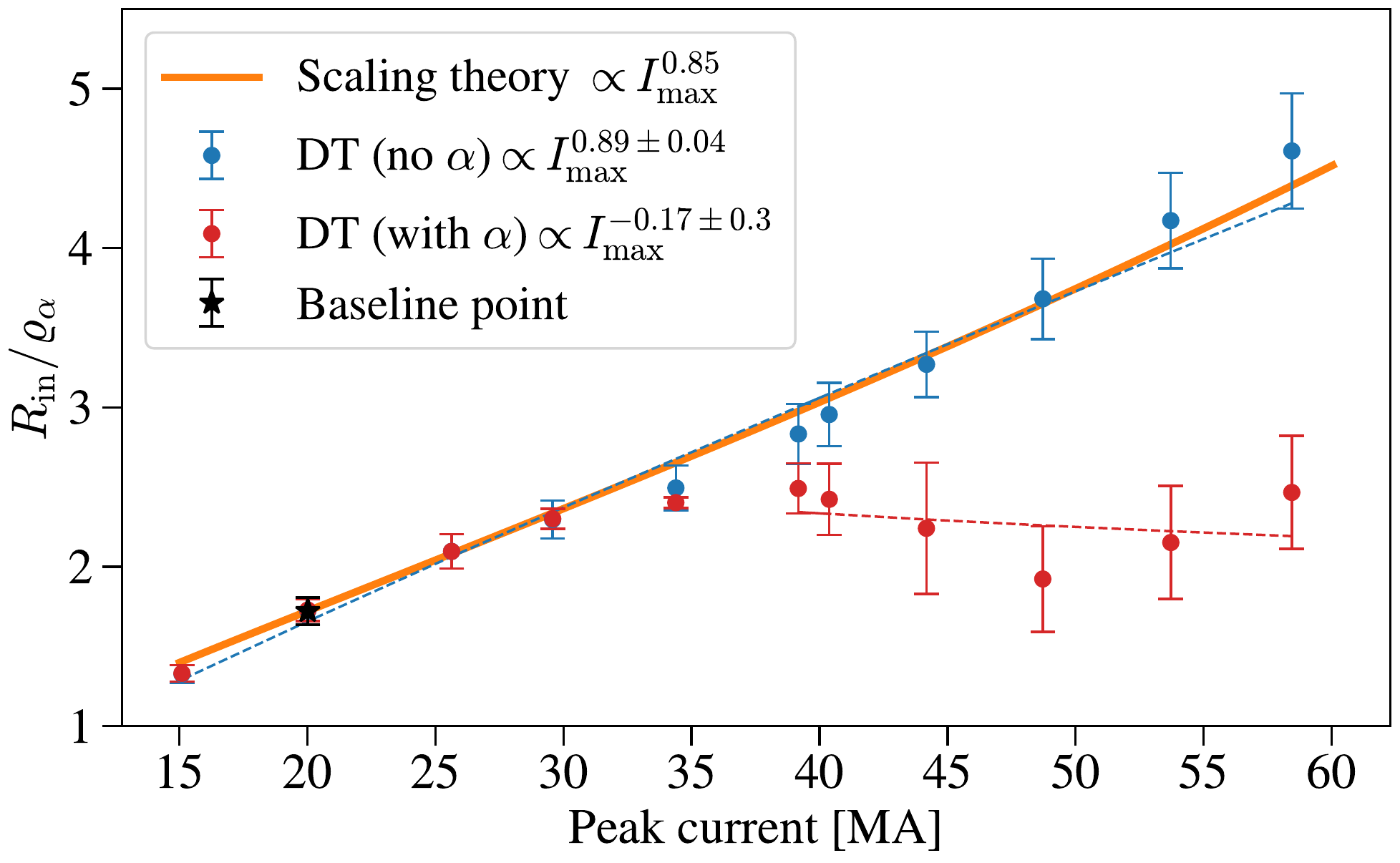}
	\caption{{\color{black}Burn-history averaged ratio of the fuel column radius $R_{\rm in}$ and the gyroradius $\varrho_\alpha$ of 3.5-MeV $\alpha$ particles.  Error bars denote the burn-weighted standard deviation associated to temporal variations near peak burn.}}
	\label{fig:BzR}
\end{figure}

\begin{figure}
	\includegraphics[scale=.43]{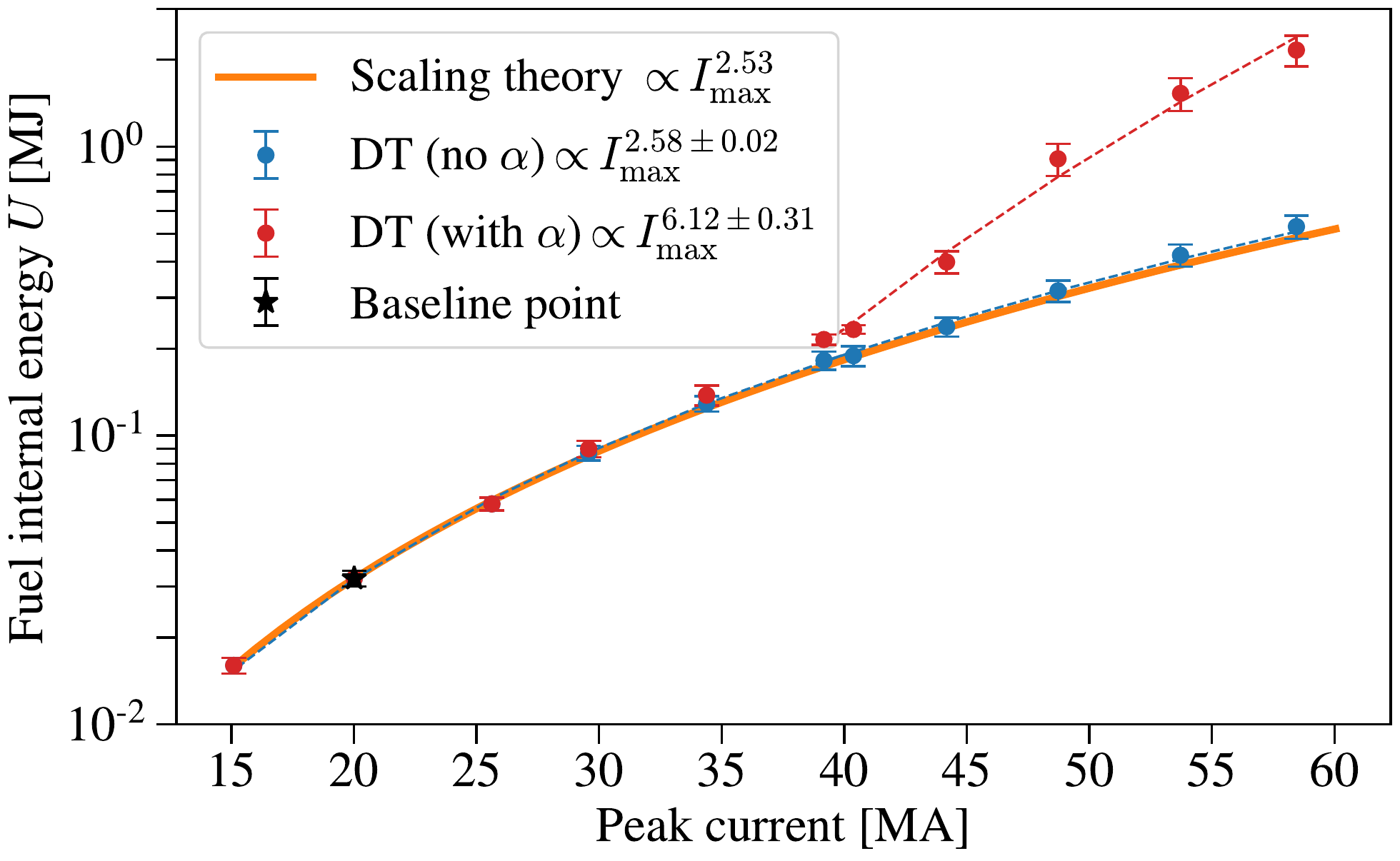}
	\caption{Fuel internal energy averaged over the neutron yield-rate history.  Error bars denote the burn-weighted standard deviation associated to temporal variations near peak burn.}
	\label{fig:U}
\end{figure}
 
{\color{black}In magneto inertial fusion, an important quantity of interest is the ratio of the fuel column radius $R_{\rm in}$ and the gyroradius $\varrho_\alpha$ of 3.5-MeV $\alpha$ particles.  $R_{\rm in}/\varrho_\alpha$ is proportional to the magnetic-field--radius product:
\begin{equation}
	\frac{R_{\rm in}}{\varrho_\alpha}
		\doteq \frac{[B_z(\mathrm{T})] \cdot [R_{\rm in} (\mathrm{cm})]}{26.5}.
\end{equation}  
When $R_{\rm in}/\varrho_\alpha \gg 1$, $\alpha$ particles are well magnetically confined within the MagLIF stagnation column.\cite{Basko:2002dv}  The magnetic-field--radius product $\smash{\langle B_zR_{\rm in} \rangle}$ is a common performance metric inferred in present-day MagLIF implosions.\cite{Schmit:2014fg,Knapp:2015kc,Lewis:2021kz}  The scaling rule for $R_{\rm in}/\varrho_\alpha$ is
\begin{equation}
	\frac{\langle R_{\rm in}/\varrho_\alpha \rangle'_{\rm no\,\alpha}}{\langle R_{\rm in}/\varrho_\alpha \rangle_{\rm no\,\alpha}}
	=
	\frac{\langle B_zR_{\rm in} \rangle'_{\rm no\,\alpha}}{\langle B_zR_{\rm in} \rangle_{\rm no\,\alpha}}
		\simeq	\frac{B_{z,0}'}{B_{z,0}}\frac{R_{\rm in,0}'}{R_{\rm in,0}}
		\simeq	\left(  \frac{\I'}{\I} \right)^{0.85}.
	\label{eq:stagnation:BzR}
\end{equation}
%0.647+0.206
Figure~\ref{fig:BzR} compares the scaling law \eq{eq:stagnation:BzR} to the simulation results. In simulations, $\langle B_zR_{\rm in} \rangle$ is calculated using
\begin{equation}
	\langle B_zR_{\rm in} \rangle(t) 
		\doteq \frac{1}{\pi \langle R_{\rm in} \rangle h } 
				\int_{V_{\rm fuel}} B_z \, \mathrm{d}V,
\end{equation}
where $\langle R_{\rm in} \rangle(t)$ is the average inner radius along the axial length of the liner and $h$ is the liner height.  As shown in \Fig{fig:BzR}, the theoretical scaling curve shows good agreement with the no-$\alpha$ simulation results.  The magnetic confinement of $\alpha$ particles increases when scaling to higher currents.  For the simulation results with $\alpha$ heating, $\langle R_{\rm in}/\varrho_\alpha \rangle'_{\rm \alpha}$ seems to reach a  threshold value of 2.5.  The value of $\langle R_{\rm in}/\varrho_\alpha \rangle'_{\rm \alpha}$ is limited by the decrease of the convergence ratio shown in \Fig{fig:CR} and by the increase of  magnetic-flux losses due to higher fuel temperatures which increase Nernst advection.  [As discussed in Paper~I, the dimensionless parameter characterizing magnetic-flux losses is proportional to the dimensionless parameter $\Upsilon_c$, whose scaling is discussed in \Fig{fig:Loss}~(right).]}

The present similarity-scaling theory allows us to estimate scaling laws for other volume-integrated quantities such as the fuel internal energy and the kinetic energy of the liner.  Using the former as an example, the no-$\alpha$ fuel internal energy $U$ evaluated at peak burn should satisfy
\begin{equation}
	\frac{U_{\rm no\, \alpha}'}{U_{\rm no\, \alpha}}
		\simeq	\frac{E_{\rm preheat}'}{E_{\rm preheat}}
		=	\left( \frac{\I'}{\I} \right)^{2.53}.
\end{equation}
As shown in \Fig{fig:U}, the theoretical scaling curve shows agreement with the averaged no-$\alpha$ fuel internal energy near peak burn.  This confirms that, when similarity scaling MagLIF loads, the no-$\alpha$ fuel internal energy scales linearly with the preheat energy.  From \Fig{fig:U}, we also note that, when scaling from 20-MA to 60-MA peak current, a roughly 16-fold increase is expected in the no-$\alpha$ internal energy of the fuel.  As previously shown, for peak currents exceeding 40 MA, calculations with $\alpha$ heating included show a sharp increase in the fuel internal energy.  

\begin{figure}
	\includegraphics[scale=.43]{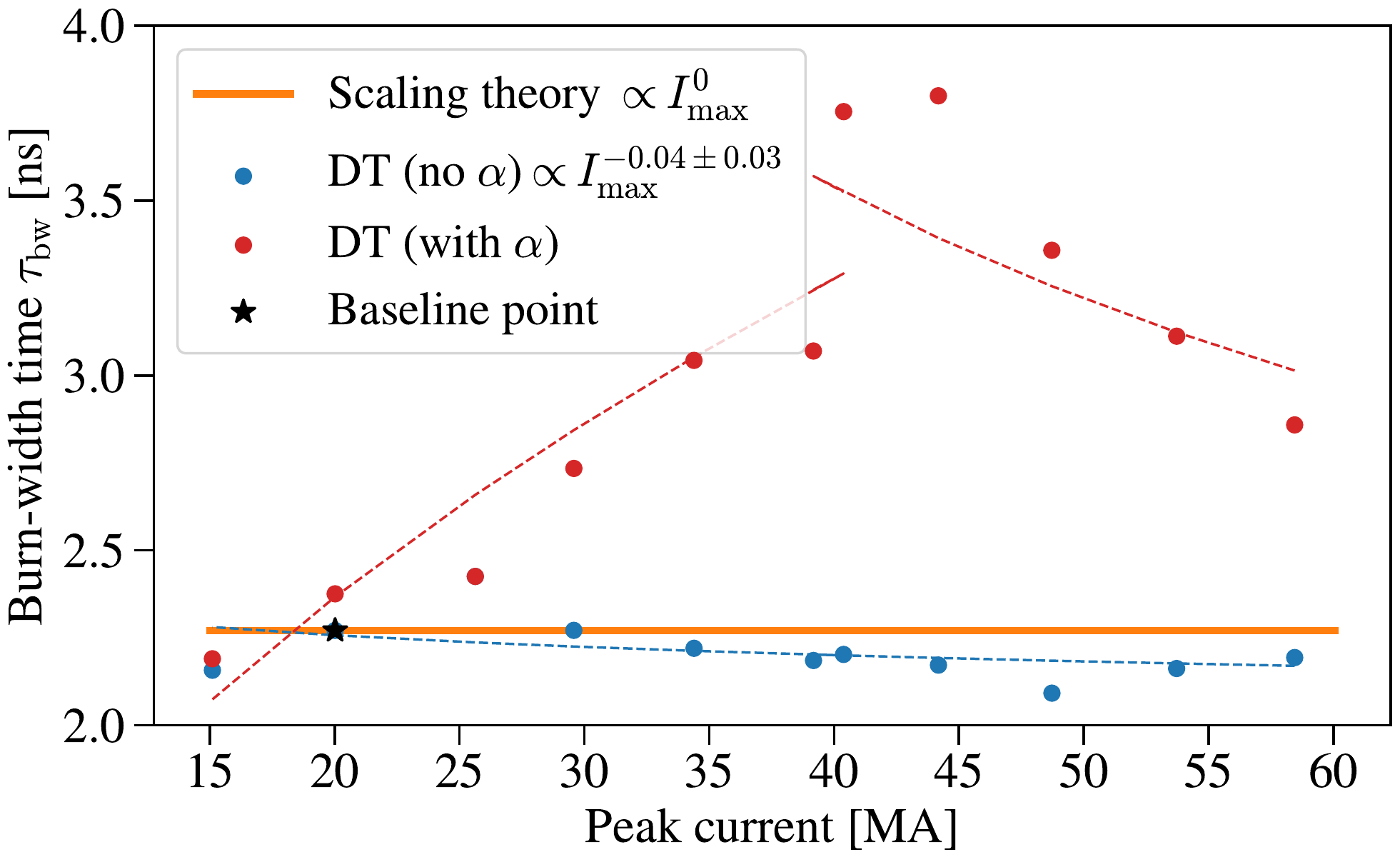}
	\caption{Burn width $\tau_{\rm bw}$ calculated by measuring the full-width, half-maximum of the neutron yield-rate time traces.  $\tau_{\rm bw}$ shows 10$\%$ variability for the simulations without $\alpha$ heating.  For the calculations with $\alpha$ heating, we obtain a power law fit $\tau_{\rm bw} \propto I_{\rm max}^{0.47\pm0.09}$ for $I_{\rm max} \leq 40$~MA and $\tau_{\rm bw} \propto I_{\rm max}^{-0.42\pm0.29}$ for $I_{\rm max} \geq 40$~MA.}
	\label{fig:tau}
\end{figure}

%%%%%%%%%%%%%%%%%%%%%%%%%%%%%%%%%%%%%%%%%%%%%%%%%
%%%%%%%%%%%%%%%%%%%%%%%%%%%%%%%%%%%%%%%%%%%%%%%%%
%%%%%%%%%%%%%%%%%%%%%%%%%%%%%%%%%%%%%%%%%%%%%%%%%
\section{Burn width and conservation of relative losses}
\label{sec:loss}

\begin{figure}
	\includegraphics[scale=.43]{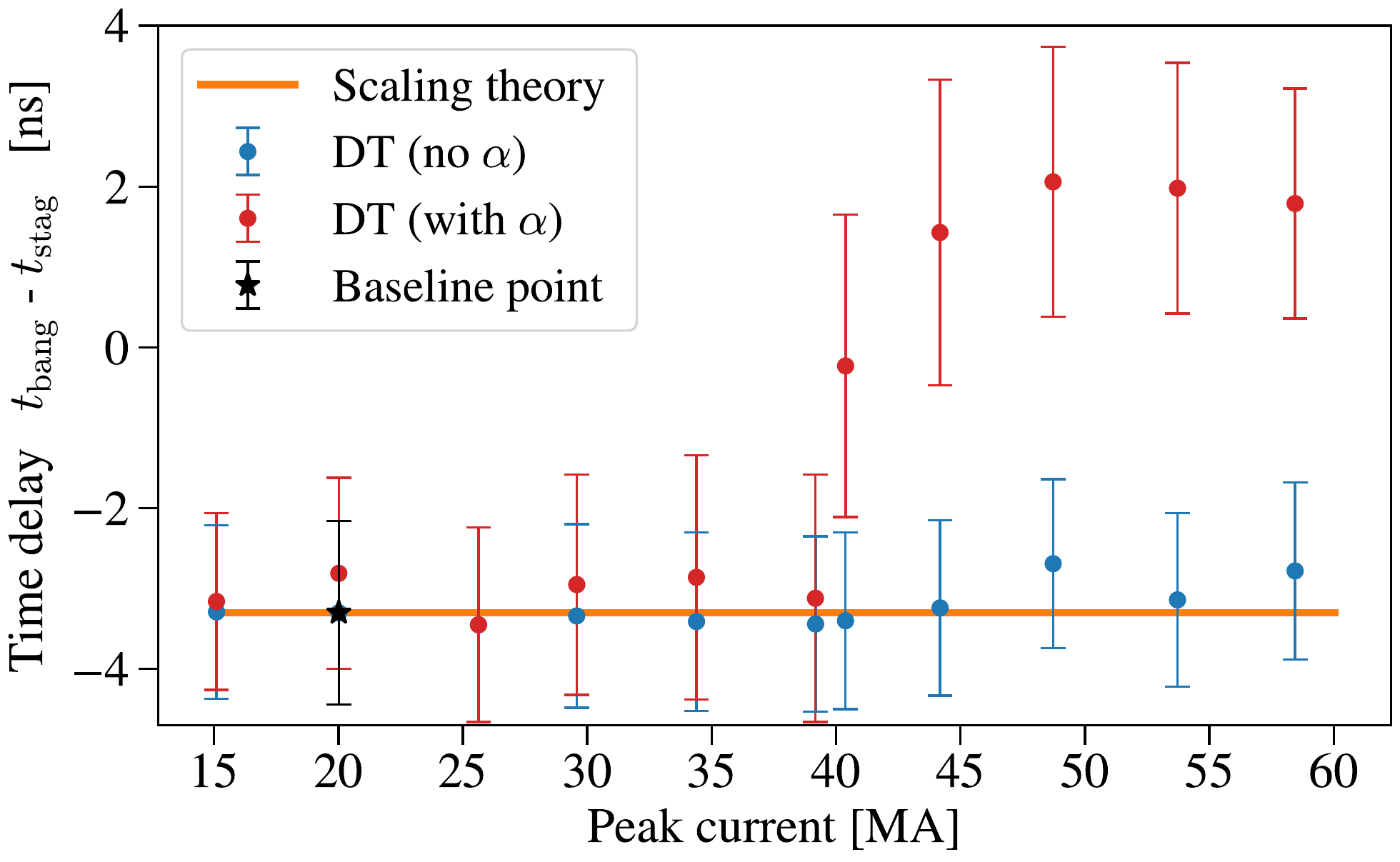}
	\caption{Time delay between the peak-burn time $t_{\rm bang}$ and the peak-compression time $t_{\rm stag}$.  For reference purposes, the error bars denote the full-width, half-maximum burn width time $\tau_{\rm bw}$ of the neutron-production events.}
	\label{fig:tdelay}
\end{figure}

In the scaling approach presented in this paper, all timescales in a MagLIF implosion are expected to remain constant.  We further test the invariance of timescales by comparing the burn-width time $\tau_{\rm bw}$ of the simulated yield rates.  As shown in \Fig{fig:tau}, $\tau_{\rm bw}\simeq 2.3$~ns is closely conserved for the simulations without $\alpha$ heating.  For the calculations with $\alpha$ heating, $\tau_{\rm bw}$ increases when increasing the peak current up to 40 MA.  Around 40-MA peak current, the burn width reaches a maximum value of $\tau_{\rm bw}\simeq3.8$~ns but then decreases at higher currents.  The physical explanation for this non-monotonic behavior is the following.  For $I_{\rm max} \leq 40$~MA, the time $t_{\rm bang}$ of peak burn occurs \emph{before} the time $t_{\rm stag}$ of peak compression of the fuel (see \Fig{fig:tdelay}).  In this regime, the pdV work rate $P_{\rm pdV}$ done on the fuel [see Eq.~(53) of Paper~I] is positive since the fuel is still imploding.  In this regime, the burn width $\tau_{\rm bw}$ increases as $\alpha$-heating effects become more dominant since fuel pressures and temperatures are increasing with current (see \Figs{fig:pion} and \ref{fig:tion}).  As shown in \Fig{fig:tdelay}, peak burn occurs \emph{after} peak compression for $I_{\rm max}>40$~MA.  Although $\alpha$ heating continues to be more prominent as current is increased, $P_{\rm pdV}$ is negative during peak burn since the fuel column is expanding.  Therefore, $P_{\rm pdV}$ acts as an energy sink that causes the burn width to decrease.  Interestingly, the change in regime shown in \Fig{fig:tdelay} correlates with the change of behavior in the scaling of the stagnation quantities discussed in \Sec{sec:stagnation}.

\begin{figure*}
	\includegraphics[scale=.43]{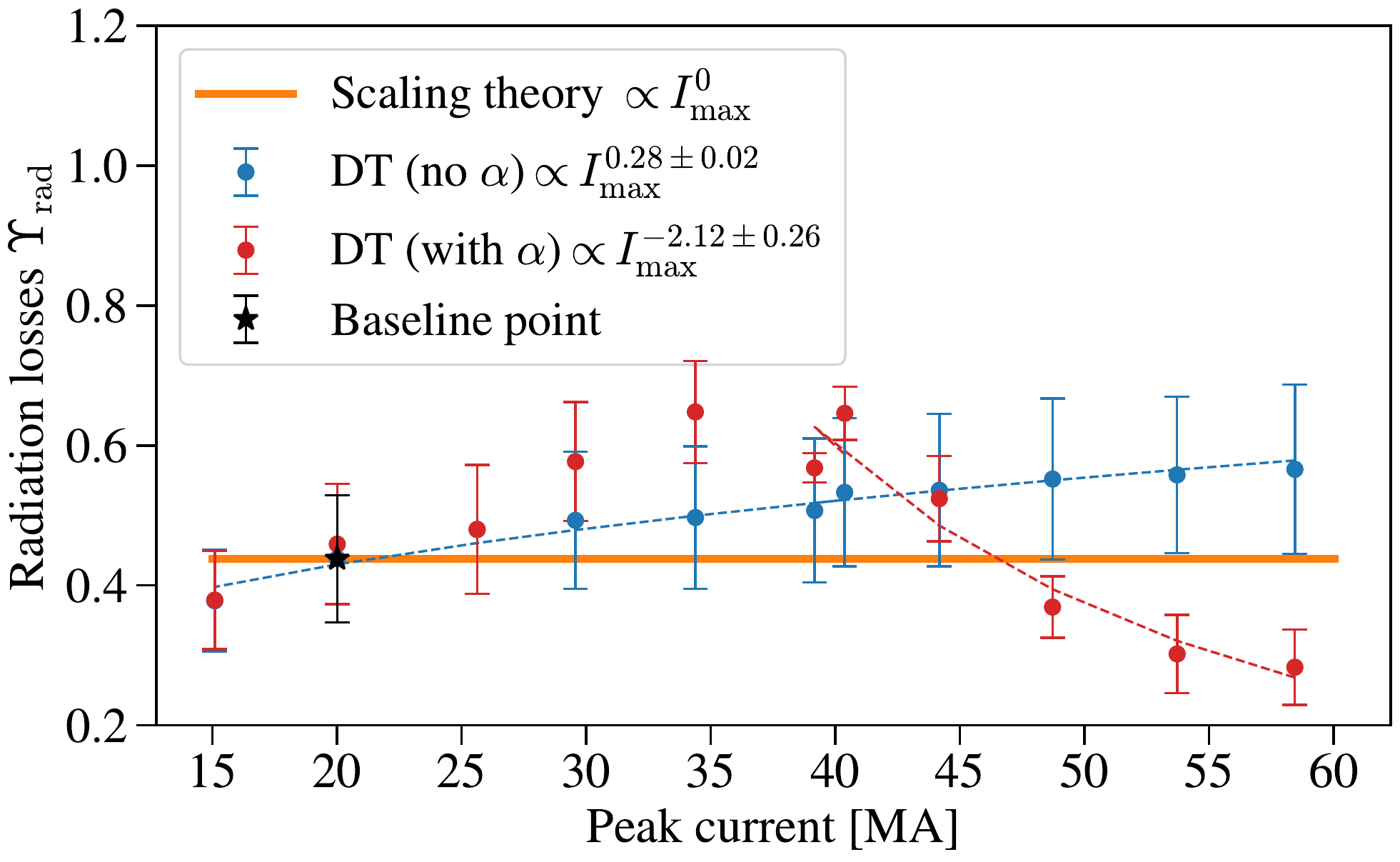}
	\hspace{0.5cm}
	\includegraphics[scale=.43]{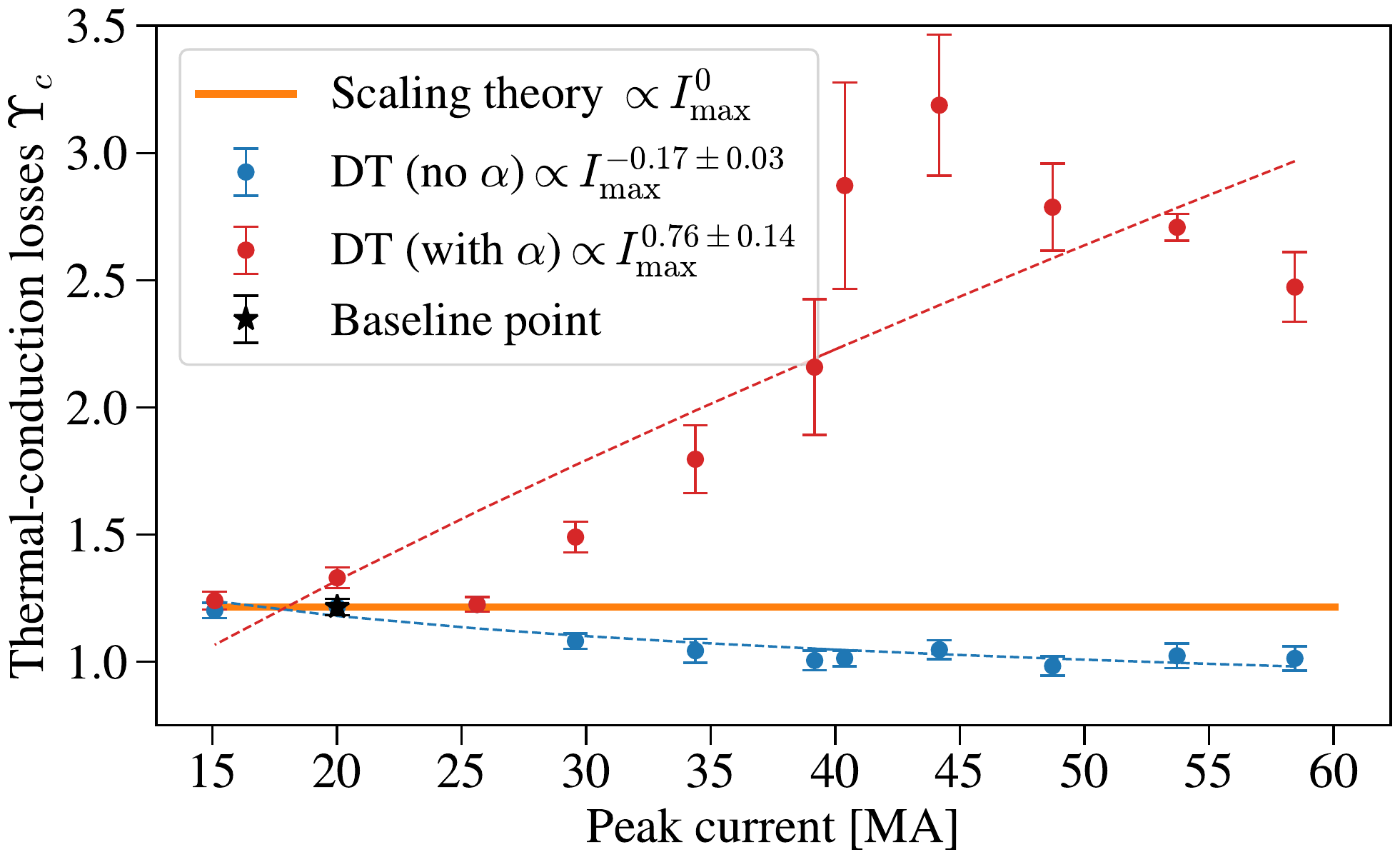}
	\caption{Left: Relative radiation energy losses as characterized by the parameter $\Upsilon_{\rm rad}$ in \Eq{eq:loss:Upsilon_rad}. Right: Relative ion-conduction energy losses as characterized by the parameter $\Upsilon_c$ in \Eq{eq:loss:Upsilon_c}.  These quantities are evaluated by substituting burn-weighted plasma parameters using \Eq{eq:stagnation:average} representing the plasma conditions near the hot plasma column.  Error bars denote the burn-weighted standard deviation associated with temporal variations near peak burn.}
	\label{fig:Loss}
\end{figure*}

\begin{figure}
	\includegraphics[scale=.43]{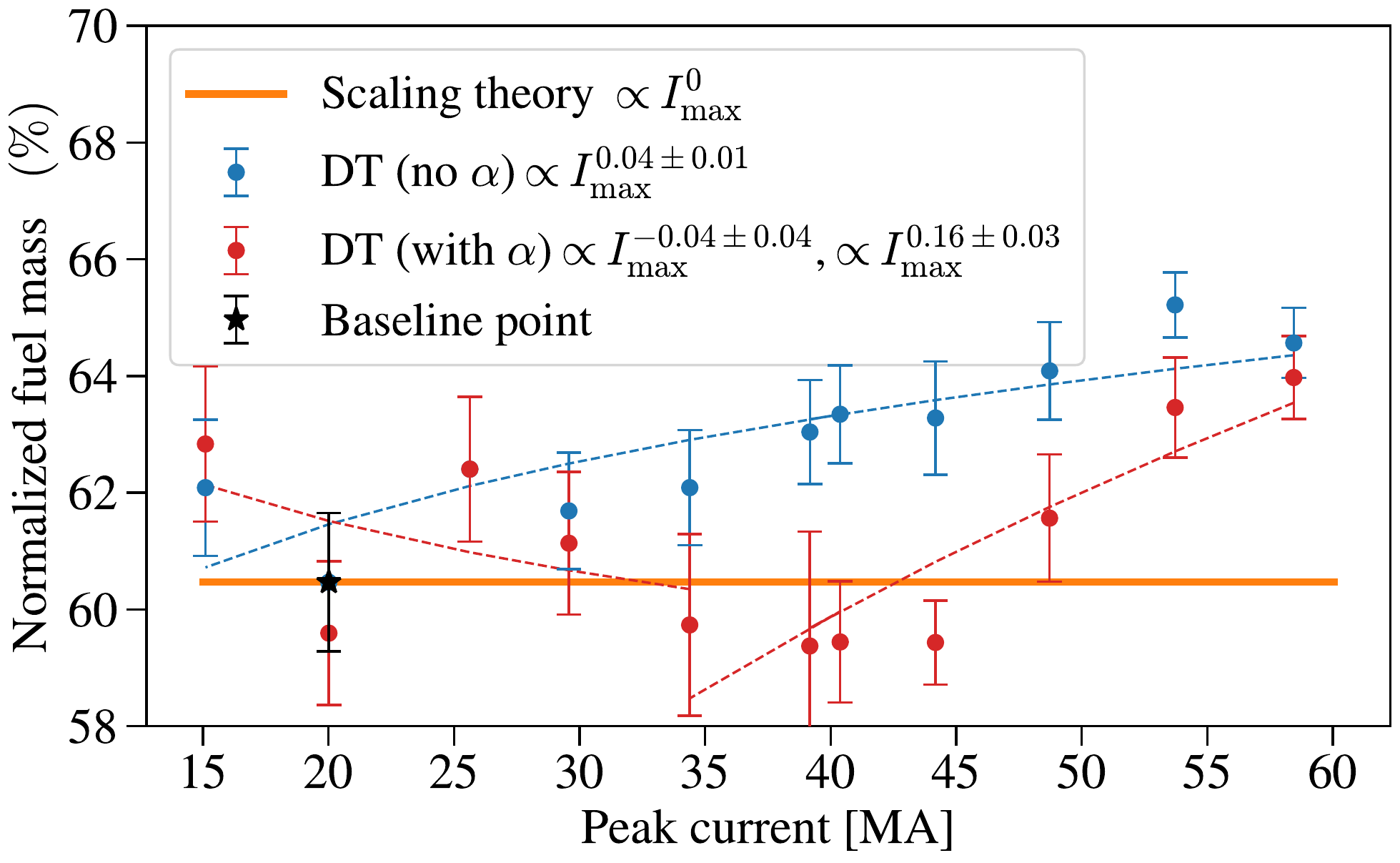}
	\caption{Burn-history averaged normalized fuel mass $\bar{m}_{\rm fuel}(t) \doteq m_{\rm fuel}(t)/m_{\rm fuel}(0)$.  Error bars denote the burn-weighted standard deviation associated to temporal variations near peak burn.}
	\label{fig:NormMass}
\end{figure}

There are two main factors that determine the burn-width time.  The first is the implosion dynamics of the liner, which we have shown to be conserved in \Sec{sec:implosion}.  The second is the energy-gain and energy-loss mechanisms.  In the no-$\alpha$ heating calculations, only energy-loss mechanisms are present.  As discussed in \Sec{sec:scaling}, the scaling prescriptions for the initial fuel density and for the axial magnetic field are designed to conserve relative radiation losses and {\color{black}thermal-conduction} losses.  Near peak burn, the relative effects of these processes can be measured by the dimensionless parameters $\Upsilon_{\rm rad}$ and ${\color{black}\Upsilon_c}$ given in Sec.~V of Paper~I:
\begin{align}
	\Upsilon_{\rm rad}	
			& \doteq	{\color{black}0.67}~
						\frac{\left[\rho ({\rm g/cm^3})\right] 
									\cdot [\tau_{\rm bw}({\rm ns} )] }
						{{\color{black}A \cdot }[T (\mathrm{keV})]^{1/2}  },
		\label{eq:loss:Upsilon_rad} \\
 	 \Upsilon_c	
			& \doteq	{\color{black}0.04 ~
						\frac{ [T (\mathrm{keV})] \cdot  [t_{\rm bw}({\rm ns} )]}
						{ \left[ B_z ({\rm T})\right] 	\cdot [R_{\rm in}({\rm cm})]^2 },}
		\label{eq:loss:Upsilon_c}
\end{align}
{\color{black}where $A=2.5$ for DT fuel.}  In \Fig{fig:Loss}, the parameters $\Upsilon_{\rm rad}$ and ${\color{black}\Upsilon_c}$ are evaluated using the calculated stagnation conditions characterizing the hot fuel column.  For the no-$\alpha$ calculations, both parameters $\Upsilon_{\rm rad}$ and ${\color{black}\Upsilon_c}$ do not deviate significantly from their baseline values.  $\Upsilon_{\rm rad,no\,\alpha}$ tends to deviate to larger values than the nominal, while ${\color{black}\Upsilon_{\rm c,no\,\alpha}}$ tends to shift towards smaller values.  This behavior is explained by the deviation observed in \Fig{fig:tion} for the ion temperature.  Since the plasma pressures and the inner convergence ratios follow the expected scaling trends, the slightly lower power law observed in the no-$\alpha$ simulations in \Fig{fig:tion} suggests that the fuel density increases slightly faster than expected causing radiation losses to become slightly stronger.  In a similar manner, the weaker scaling in ion temperature shown in \Fig{fig:tion} explains the decrease in ${\color{black}\Upsilon_{\rm c,no\,\alpha}}$ shown in \Fig{fig:Loss}.  {\color{black}Interestingly, the parameter $\Upsilon_c$ with $\alpha$ heating included increases considerably at larger peak currents.  The increase in $\Upsilon_c$ is due to several reasons: the higher fuel temperatures and lower magnetic fields at stagnation and the longer burn width time $\tau_{\rm bw}$ as $\alpha$ heating becomes more important.}

In order to conserve relative end-flow energy losses and fuel-mass losses when scaling in peak current, the axial height of MagLIF loads is varied according to the scaling prescription in \Eq{eq:numerical:rho}.  To measure the {\color{black}effectiveness} of this scaling rule, we tallied the total fuel mass inventory $m_{\rm fuel}(t)$ located within the imploding region of a MagLIF load.  Based on similarity-scaling arguments, it is expected that the normalized fuel mass $\bar{m}_{\rm fuel}(t) \doteq m_{\rm fuel}(t)/m_{\rm fuel}(0)$ should remain invariant when scaling across currents.  Figure~\ref{fig:NormMass} shows the normalized fuel mass evaluated near peak burn.  Our calculations with and without $\alpha$ heating suggest that about 60\%--66\% of the initial fuel inventory remains in the imploding region up to the moment of peak burn.  Figure~\ref{fig:NormMass} shows a variation below 10\% in the normalized fuel-mass inventory, which confirms that the scaling law \eq{eq:numerical:rho} for the load height is overall conserving relative end losses.  However, we must note that \Eq{eq:numerical:rho} dictates a relatively large increase in the axial length of a MagLIF liner (as shown in \Figs{fig:numerical:parameters} and \ref{fig:liners}).  This is inconvenient due to the additional initial inductance associated with longer liners, which makes it more difficult to  deliver higher peak currents with a given pulsed-power generator.  It may be possible to reduce the scaling exponent of the load height by modifying the scaling prescriptions of the radial dimensions of the laser-entrance-hole window and the cushions.  In this work, these parameters were scaled linearly with the initial inner radius of the liner [see \Eq{eq:scaling:Rcushion}].  However, smaller end openings could reduce end losses and thus decrease the axial length of the similarity-scaled MagLIF loads.  {\color{black}A more thorough investigation of end losses and methods to mitigate them} will be left for future work.

%%%%%%%%%%%%%%%%%%%%%%%%%%%%%%%%%%%%%%%%%%%%%%%%%
%%%%%%%%%%%%%%%%%%%%%%%%%%%%%%%%%%%%%%%%%%%%%%%%%
%%%%%%%%%%%%%%%%%%%%%%%%%%%%%%%%%%%%%%%%%%%%%%%%%
\section{Scaling of MagLIF performance}
\label{sec:performance}

The similarity-scaling framework in \Sec{sec:scaling} leads to good agreement between the theory and simulations for the estimated plasma stagnation conditions and the burn-width time $\tau_{\rm bw}$.  Now, we compare the metrics for the expected performance of the similarity-scaled MagLIF loads.  The fusion yield follows the scaling of the characteristic yield number $Y_{\rm ref}$ introduced in Paper I.  The no-$\alpha$ yield $Y_{\rm no \, \alpha}$ obeys the following scaling rule:
\begin{equation}
	\frac{Y_{\rm no \, \alpha}'}{Y_{\rm no \, \alpha}}
		\simeq 	\left( \frac{\rho_0'}{\rho_0} \right)^2
			 \left( \frac{T_{\rm preheat}'}{T_{\rm preheat}} \right)^{3.77}
			 \left( \frac{R_{\rm in,0}'}{R_{\rm in,0}} \right)^2
			 \frac{h'}{h}.
\end{equation}  
%2*0.529+3.77*1.06+2*0.206+0.529=5.99
Here we used the power-law fit in Eq.~(109) of Paper~I for the DT fusion reactivity so that $\langle \sigma v \rangle_{\rm DT} \propto T^{3.77}$ which is valid within the 2--8~keV range shown in \Fig{fig:tion} for the no-$\alpha$ temperatures.  Substituting \Eqs{eq:numerical:R}, \eq{eq:numerical:rho}, and \eq{eq:stagnation:tion2}, we obtain the scaling rule for the no-$\alpha$ fusion yield:
\begin{equation}
	\frac{Y_{\rm no \, \alpha}'}{Y_{\rm no \, \alpha}}
			\simeq	\left( \frac{\I'}{\I} \right)^{5.99}.
	\label{eq:performance:yield}
\end{equation}

\begin{figure}
	\includegraphics[scale=.43]{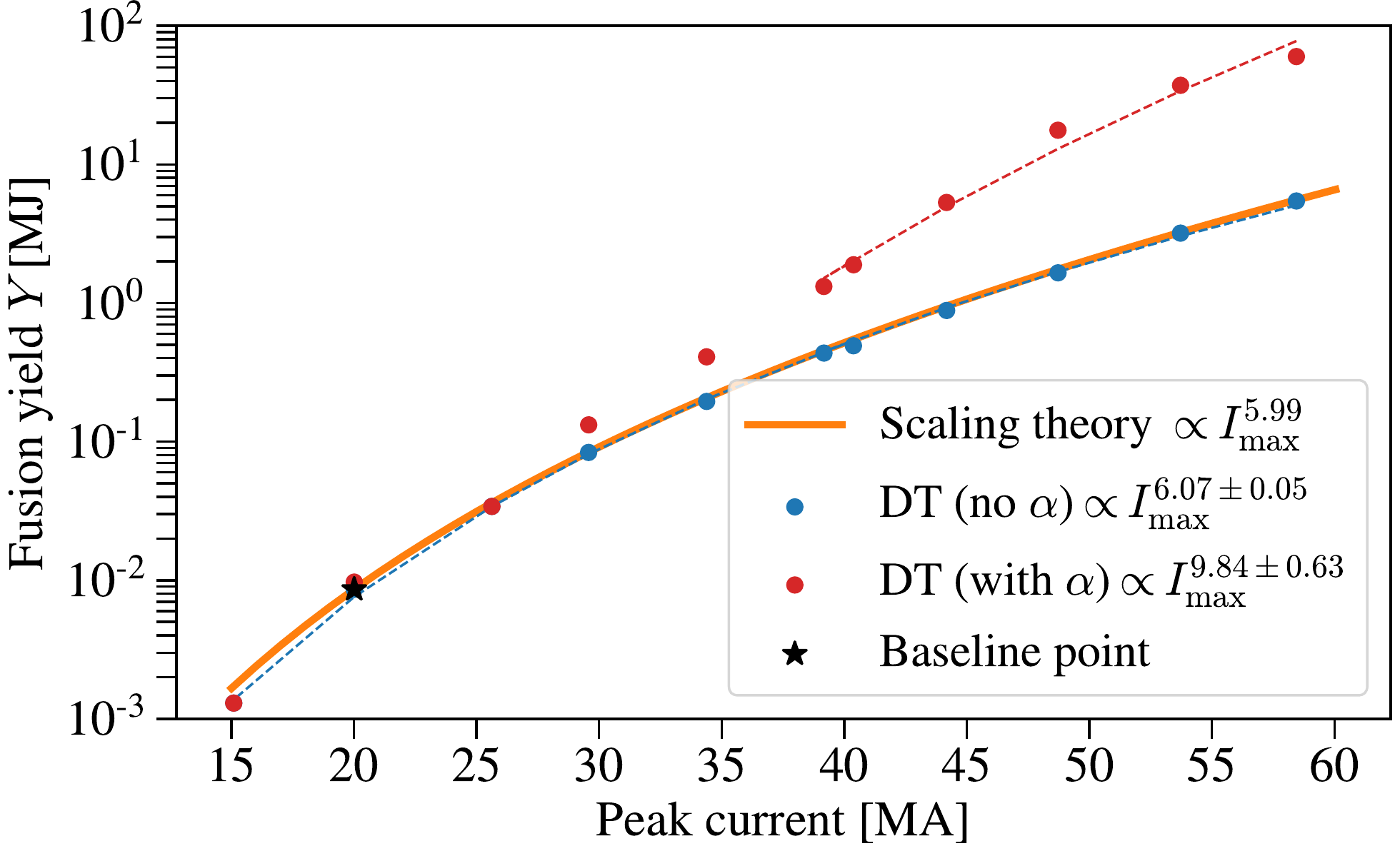}
	\caption{Fusion yield of similarity-scaled MagLIF loads.  Red and blue points denote simulation results with and without $\alpha$ heating, respectively.   Dashed lines are power-law fits to the simulation data.  The legend shows the fitted scaling exponents.  The orange curve is the scaling law in \Eq{eq:performance:yield}.}
	\label{fig:yield}
\end{figure}

The scaling of the yield per-unit-length $\smash{\widehat{Y}_{\rm no \, \alpha}}$ is found by removing the factor related to the liner height:
\begin{equation}
	\frac{\widehat{Y}_{\rm no \, \alpha}'}{\widehat{Y}_{\rm no \, \alpha}}
			\simeq	\left( \frac{\I'}{\I} \right)^{5.47}.
	\label{eq:performance:yieldhat}
\end{equation}
%2*0.529+3.77*1.06+2*0.206=5.47
This scaling law is more favorable than the often quoted $\smash{\widehat{Y}_{\rm no \, \alpha} \propto \I^4}$ scaling for z-pinch devices.\cite{Velikovich:2007hq}  This occurs for two reasons.  First, as a consequence of the scaling constraints on the preheat energy and on the liner inner radius (which scales relatively weakly with $\I$ to mitigate MRT feedthrough), the relatively more compact scaled fuel volumes are predicted to achieve higher fuel pressures and temperatures.  Second, the initial fuel density is scaled sublinearly with respect to current to maintain the relative effects of radiation losses.  This leads to the almost linear increase in ion temperatures near stagnation shown in \Fig{fig:tion}, which in turn increases the DT neutron reactivity.

Figure \ref{fig:yield} shows the fusion yields for the similarity-scaled MagLIF loads and compares them to the analytical estimate in \Eq{eq:performance:yield}.  The theory and the simulation results without $\alpha$ heating show excellent agreement.  It is worth noting that  the no-$\alpha$ yield varies by nearly three orders of magnitude when varying the peak current from 15~MA to 60~MA.  {\color{black}In terms of absolute yield numbers, the no-$\alpha$ fusion yield for the anchor load driven at 20 MA is $8.6$~kJ, and the theoretically expected no-$\alpha$ yield for the 60-MA load is $0.0086\cdot(60/20)^{5.99} \simeq 6.2$~MJ, which agrees with the $5.4$~MJ no-$\alpha$ yield at 60 MA.  Interestingly, the calculations with $\alpha$ heating show that similarity-scaled MagLIF loads can self-heat at higher currents and can lead to yields of roughly $Y\simeq60$~MJ at the 60-MA level.}

\begin{figure}
	\includegraphics[scale=.43]{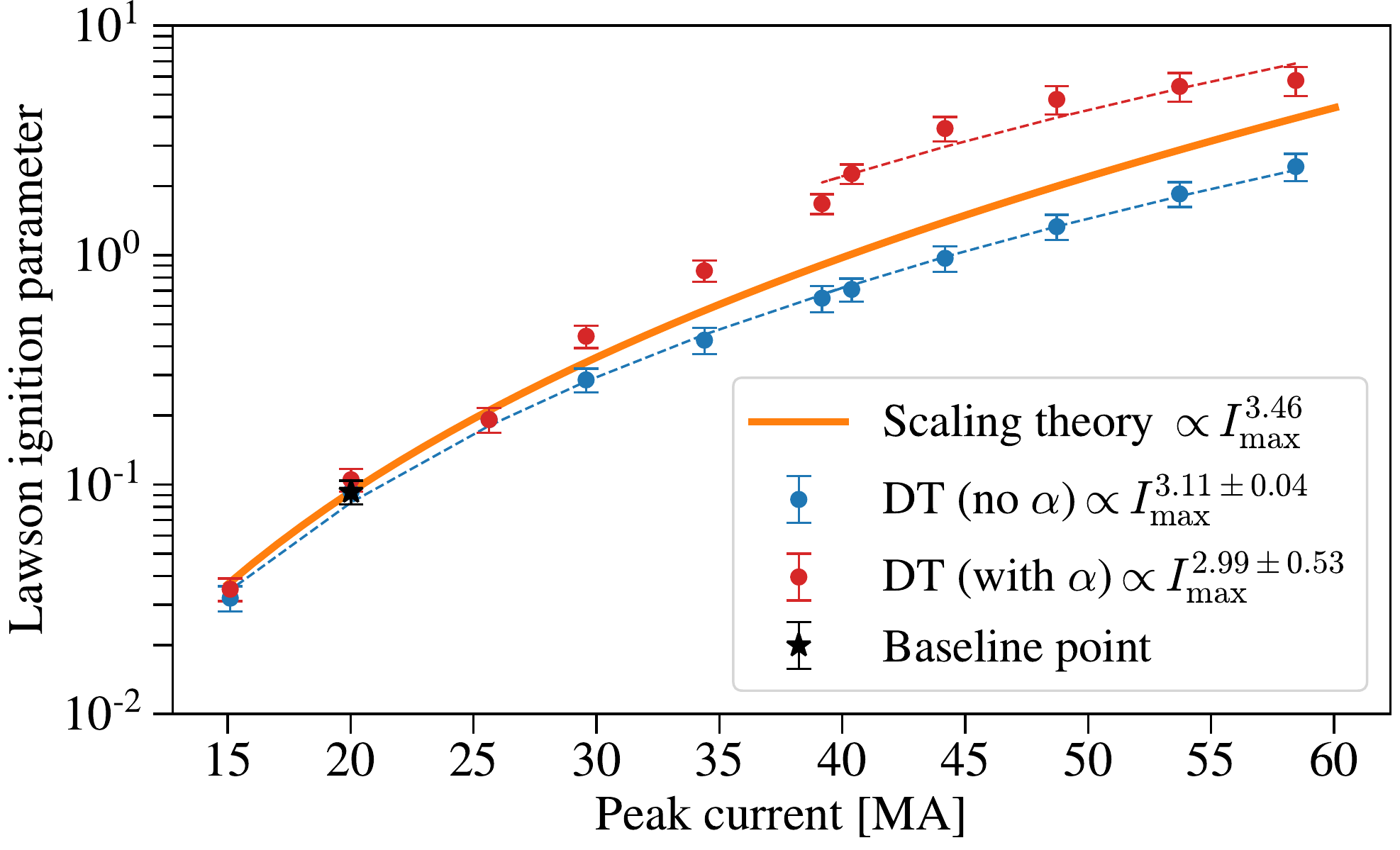}
	\caption{Burn-history averaged Lawson ignition parameter $\chi$ of the similarity-scaled MagLIF loads.  Error bars denote the burn-weighted standard deviation associated to temporal variations near peak burn of the fuel pressure and temperature entering \Eq{eq:performance:chi_convenient}.}
	\label{fig:chi}
\end{figure}

Our results for various stagnation and performance metrics suggest that scaled-up MagLIF loads can potentially reach robust $\alpha$-heating regimes.  A metric often used in the literature to measure this effect is the Lawson-ignition parameter $\chi$.\cite{Betti:2010fc}  Following Paper I, this quantity can be written as follows:
\begin{equation}
	\chi = 0.09 \, [p_{\rm fuel}({\rm Gbar})] \cdot [\tau_{\rm bw}({\rm ns})] \,
	\frac{[ \langle \sigma v \rangle (10^{-18} {\rm cm^3/s})] }{[T ({\rm keV}) ]^2},
	\label{eq:performance:chi_convenient}
\end{equation}
where we have dropped the term $\eta_\alpha$ measuring the fraction of trapped $\alpha$ particles.    The no-$\alpha$ Lawson parameter $\chi_{\rm no\, \alpha}$ obeys the following scaling rule:
\begin{equation}
	\frac{\chi_{\rm no\, \alpha}'}{\chi_{\rm no\, \alpha}}
		\simeq \frac{p_{\rm preheat}'}{p_{\rm preheat}} 
					\left( \frac{T_{\rm preheat}'}{T_{\rm preheat}}	\right)^{1.77}
		\simeq \left( \frac{\I'}{\I} \right)^{3.46}.
\end{equation}
%1.59+1.77*1.06=3.46
Figure \ref{fig:chi} compares the theoretical scaling curve for the $\chi$ parameter and the burn-history averaged $\chi$ values obtained from simulations.  When comparing to the no-$\alpha$ results, we find that the theoretical scaling rule $\chi_{\rm no\, \alpha} \propto \I^{3.46}$ over-predicts the simulated scaling curve which shows a ${\color{black}\chi_{\rm no\, \alpha} \propto \I^{3.11}}$.  The discrepancy between the theoretical and the fitted scaling exponents is explained by the observed deviation in the scaling in the fuel temperature shown in \Fig{fig:tion}.  It is worth noting that MagLIF experiments at the 20-MA scale have demonstrated $\chi_{\rm no\, \alpha}\simeq 0.1$.\cite{foot:Knapp}  Therefore, MagLIF on Z is far from a robust $\alpha$-heating regime.  This is expected since the Z facility does not have enough energy to reach such thermonuclear conditions.  However, when similarity scaling the MagLIF platform to 60 MA, we find that $\chi_{\rm no\, \alpha}\simeq 2.4$, which surpasses the ignition threshold of unity.  The calculated $\chi$ parameter for the simulations with $\alpha$ heating surpass the $\chi=1$ threshold around 45-MA peak current.  This correlates well with the regime changes observed for the burn-averaged fuel parameters, such as the fuel pressure and temperature.  Although $\chi_\alpha$ increases sharply in the 35-45 MA range, the fitted scaling curve at higher currents has a similar exponent as that of the no-$\alpha$ calculations.  When considering the fitted power-laws for $p_{\rm fuel, \alpha}$ and $T_{\rm \alpha}$, this behavior for $\chi_\alpha$ can be partially attributed to the decrease in the burn-width time discussed in \Fig{fig:tau} and to a smaller power-law exponent for the DT reactivity for high temperatures between 5-20 keV $(\langle \sigma v\rangle_{\rm DT} \propto T^{2.4})$.

{\color{black}As a reminder,} the simulation results in this paper are obtained from 2D ``clean" simulations that do not consider impurity mixing nor initial seeding of the MRT instability in the outer surface of the liner.  Therefore, these results are inherently optimistic and over-predict the fusion yields that would be observed in experiment.  Nevertheless, the main takeaways are (1)~we propose a new paradigm for scaling MagLIF loads to higher currents, (2)~we have tested the theory against several metrics describing the implosion dynamics, stagnation conditions, and performance, and (3)~the theory and the no-$\alpha$ simulation results show agreement.  These results increase our confidence of using this scaling paradigm to explore the performance of scaled-up MagLIF configurations.  For future work, we shall use this scaling framework to scale MagLIF loads while including mixing and instability effects in order to better assess the potential of MagLIF to reach high yields at higher currents.

%%%%%%%%%%%%%%%%%%%%%%%%%%%%%%%%%%%%%%%%%%%%%%%%%
%%%%%%%%%%%%%%%%%%%%%%%%%%%%%%%%%%%%%%%%%%%%%%%%%
%%%%%%%%%%%%%%%%%%%%%%%%%%%%%%%%%%%%%%%%%%%%%%%%%
\section{Conclusions}
\label{sec:conclusions}

The MagLIF platform is a magneto-inertial-fusion concept studied on the Z Pulsed Power Facility.\cite{Gomez:2014eta,Knapp:2019gf,Gomez:2019bg,Gomez:2020cd,YagerElorriaga:2022cp,Sinars:2020bv}  Given the relative success of this platform, we proposed a novel method to scale MagLIF to higher currents in order to reach higher yields.  Our method is based on similarity scaling.\cite{Schmit:2020jd,foot:Ruiz_framework}  Similarity scaling attempts to preserve many of the physics regimes already known or being studied on today's Z machine with the goal of reducing unexpected outcomes on future scaled-up experiments.  In this paper, we derived scaling rules for the experimental input parameters characterizing a MagLIF load as unique functions of the characteristic current driving the implosions.  We also derived scaling rules for various no-$\alpha$ metrics describing the liner-implosion dynamics, stagnation conditions, and performance.  The scaling rules were compared against 2D radiation--magneto-hydrodynamic (rad-MHD) \textsc{hydra} simulations.\cite{Marinak:1996fs,Koning:2009}  Overall, agreement was found between the scaling theory and simulation results.  In particular, analytical and 2D ``clean" numerical calculations showed that MagLIF loads have the potential to reach {\color{black}60-MJ} yield in a 60-MA--class pulsed-power facility \textit{when similarity scaled} from MagLIF configurations presently studied on Z.\cite{Gomez:2020cd,YagerElorriaga:2022cp}

It is worth mentioning that the projected yields in this work are lower than the $\sim440$-MJ yield at $\sim 65$~MA peak current calculated from the optimized-scaling studies in \Refa{Slutz:2016cf}.  The reduction in the projected yields is likely caused by three reasons. First, similarity scaling imposes strict constraints on the scaling of the MagLIF liner (specifically, the liner thickness) to maintain the implosion stability.  These constraints can limit the stagnation pressures that can be achieved at higher currents.  In contrast, the scaling studies in \Refa{Slutz:2016cf} assumed a constant initial AR=6 for all scaled liners, which are relatively thinner and more unstable at higher currents than those discussed in this paper.  Second, at high peak currents, the initial fuel densities suggested in \Refa{Slutz:2016cf} are significantly higher than those shown in this paper ($\sim$10~mg/cc compared to $\sim$4~mg/cc).  Denser fuel configurations allow for better coupling of the $\alpha$ particles with the background fuel and therefore lead to higher fusion yields.  However, laser preheat becomes a challenge with such relatively high initial fuel densities. For comparison, 15\% critical density of a 3$\omega$ laser propagating in DT plasma is 5.55~mg/cc.  Therefore, significant laser--plasma interactions (LPI) could be expected at such high densities.  Third, the circuit models used in this work and in \Refa{Slutz:2016cf} are different.  In this work, the circuit models are derived from the similarity-scaling rules and use the canonical circuit model for Z as a baseline.  The scaling rules for the circuit are designed to maintain the pulse shape of the normalized current traces (see \Fig{fig:radius_norm_vs_time}).  In contrast, the calculations shown in \Refa{Slutz:2016cf} are based on circuit models of two conceptual designs of two future petawatt-class pulsed-power accelerators (Z300 and Z800).\cite{Stygar:2015kza}  The circuit models have different assumptions on the behavior of current losses not reaching the MagLIF load.  Differences in the power delivery and current losses assumed between this work and in \Refa{Slutz:2016cf} can affect the comparisons in extrapolated performance of MagLIF loads even when considering the same peak current.

The present work can be extended in several directions.  First, it is important to identify the role of unknown physical processes (or ``hidden" variables) that can affect the scaling results presented in this paper.  Interfacial instabilities and mix within the fuel are a particular concern.  In this regard, it is important to assess how the seeding of the MRT instability by the electro-thermal instability\cite{Oreshkin:2008dy,Peterson:2012bu,Peterson:2013eh,Yu:2020gl,Awe:2021jp} behaves at higher current densities.  Future research should also focus on better the \textit{ab initio} modeling of the spontaneously generated, helical MRT modes observed in MagLIF-type implosions.\cite{Awe:2014gba,Awe:2013dt}  Important questions to answer are: (1)~how do helical modes scale with peak current and initial axial magnetic field, (2)~do helical modes lead to strong mixing of Be liner material into the fuel, and (3)~can helical MRT modes decrease the confinement of $\alpha$ particles and consequently truncate self-heating of the fuel.  These questions concerning the scaling of interfacial instabilities and mix can be addressed via dedicated experiments and 3D simulations.  Preheat delivery is another area deemed of ``higher risk" when scaling MagLIF to higher peak currents.  Regarding this topic, it will be important to experimentally demonstrate the feasibility to deliver the preheat energies required by the scaling theory, understand the effects of vortex flows seeded within the fuel by the laser preheat,\cite{Weis:2021id} and estimate the degree of laser--plasma instabilities that will occur in future, more energetic preheat configurations.  {\color{black}In this regard, \Refa{Schmit:2020jd} gives a preliminary estimates of the scaling of various LPI processes present during the MagLIF preheat stage.}  Dedicated preheat experiments \textit{at scale} are currently underway at the National Ignition Facility.\cite{foot:Pollock}  For the interested reader, it is worth mentioning that \Refa{YagerElorriaga:2022cp} reviews the present-day research status of the MagLIF effort, and \Refa{Ruiz:2022} summarizes the research needs and challenges for MagLIF with a particular emphasis towards theory, simulations, and scaling to higher peak currents.

As a second research direction, the work presented in this paper can be extended to applying the present scaling paradigm to ``ice-burning" MagLIF configurations.  In this paper, we only considered ``gas-burning" MagLIF loads, \ie loads with gaseous fuel configurations.  Numerical simulations presented in \Refs{Slutz:2012gp,Sefkow:2014ik,Slutz:2016cf} showed that ``ice-burning" MagLIF loads, \ie loads with DT ice layers on the liner inner wall, can perform significantly better at higher currents beyond 55 MA.  It would be interesting to extend the present similarity-scaling theory to this second class of MagLIF loads and assess the potential of such similarity-scaled configurations.  

As a third research direction, the similarity-scaling framework provides a roadmap to experimentally study MagLIF scaling physics on the Z facility.  This can be done by turning down the machine charge voltage and self-consistently down-scaling the MagLIF load and estimating the current delivery.  An experimental effort {\color{black}is} underway at Sandia to test the similarity-scaling theory against experiments by varying the peak current within the 14--20 MA range.  If agreement is found, results from these experiments will bolster the confidence in scaling MagLIF and will help reduce uncertainties when extrapolating MagLIF performance to higher currents.

One of the authors (D.~E.~Ruiz) was supported in part by Sandia National Laboratories (SNL) Laboratory Directed Research and Development (LDRD) Program, Project 223312.  Sandia National Laboratories is a multimission laboratory managed and operated by National Technology $\&$ Engineering Solutions of Sandia, LLC, a wholly owned subsidiary of Honeywell International Inc., for the U.S. Department of Energy's National Nuclear Security Administration under contract DE-NA0003525.  This paper describes objective technical results and analysis. Any subjective views or opinions that might be expressed in the paper do not necessarily represent the views of the U.S. Department of Energy or the United States Government.

%%%%%%%%%%%%%%%%%%%%%%%%%%%%%%%%%%%%%%%%%%%%%%%%%
%%%%%%%%%%%%%%%%%%%%%%%%%%%%%%%%%%%%%%%%%%%%%%%%%
%%%%%%%%%%%%%%%%%%%%%%%%%%%%%%%%%%%%%%%%%%%%%%%%%

\appendix

\section{Origin of the correction factor in \Eq{eq:scaling:Rout}}
\label{app:correction}

In this appendix, we provide an intuitive explanation for the origin of the correction factor introduced in \Eq{eq:scaling:Rout}.  A characteristic feature of liner implosions driven at high currents [$\mc{O}$(1-10~MA)] with relatively short rise times [$\mc{O}(100~\rm{ns})]$ is that the liner material is usually shocked.  Before shock breakout and before the liner begins to move as a whole, the outer radius of the liner is displaced inwards.  This displacement causes the electric current to travel at a smaller effective radius, which in turn, increases the magnetic pressure and the magnetic drive of the implosion.  For this reason, it is hypothesized that the liner implosion trajectories are not conserved when strictly following the scaling prescription in the first line of \Eq{eq:scaling:Rout}, which does not account for shock compression.

To obtain similarity-scaled liner implosion trajectories, we posit that, the derived scaling relation for the liner outer radius in Paper I should be applied to the radius $R_{\rm out,\star}$, which is the liner outer radius at the moment when the shock traversing the liner has broken out and the liner begins to move as a whole.  $R_{\rm out,\star}$ is related to the initial outer radius $R_{\rm out,0}$ via $R_{\rm out,\star} = R_{\rm out,0} - \Delta R$, where $\Delta R$ is the displacement of the outer surface of the liner before shock break out. Usually, $\Delta R \ll R_{\rm out,0}$.  From this consideration, the scaling prescription for the liner mass per-unit-length becomes
\begin{align}
	\left( \frac{\I'}{\I} \right)^{\frac{\gamma-1}{2\gamma-1}}
		& =	\frac{R_{\rm out,\star}'}{R_{\rm out,\star}}
		   = \frac{R_{\rm out,0}' - \Delta R'}{R_{\rm out,0} - \Delta R}.
\end{align}
As a reminder, for an arbitrary \textit{baseline} quantity $Q$, the quantity $Q'$ denotes its \textit{scaled} value.  When Taylor expanding this expression, we obtain
\begin{equation}
	\left( \frac{\I'}{\I} \right)^{\frac{\gamma-1}{2\gamma-1}}
		 \simeq  \left( \frac{R_{\rm out,0}'}{R_{\rm out,0}} \right)
						\left( 1 + \frac{\Delta R}{R_{\rm out,0}} - \frac{\Delta R'}{R_{\rm out,0}'} \right).
	\label{eq:app:Istar}
\end{equation}

The relative displacement $\Delta R/ R_{\rm out,0}$ of the liner outer radius will be generally dependent on the liner thickness, the characteristic magnetic pressure, the equation-of-state of the liner material, and the time history of the current drive.  For the similarity-scaled loads studied in this paper, the first two quantities can be considered as functions of the characteristic current $\smash{\I}$, while the last two quantities would nominally remain unchanged.  We consider small changes in $\Delta R/R_{\rm out,0}$ when scaling by the characteristic current.  We Taylor expand $\Delta R'/ R_{\rm out,0}'$ in \Eq{eq:app:Istar} and obtain
\begin{equation}
	\frac{\Delta R'}{R_{\rm out,0}'} 
		\simeq 		\frac{\Delta R}{R_{\rm out,0}}
					 + (\I' - \I) 
					 	\frac{\mathrm{d}}{\mathrm{d} \I} 
					 	\left(  \frac{\Delta R}{R_{\rm out,0}}\right) + ...
\end{equation}
Substituting this expression into \Eq{eq:app:Istar} leads to a modified scaling law for the liner outer radius:
\begin{equation}
	\frac{R_{\rm out,0}'}{R_{\rm out,0}}
		\simeq
			\left( \frac{\I'}{\I} \right)^{\frac{\gamma-1}{2\gamma-1}}
			\left[ 1 + \mc{C} \left( \frac{\I'}{\I} -1  \right) \right],
		\label{eq:app:Rout}
\end{equation}
where the term inside the square brackets is the correction factor and
\begin{equation}
	\mc{C}(\I) 
		\doteq \I
					\frac{\mathrm{d}}{\mathrm{d} \I} 
					\left(  \frac{\Delta R}{R_{\rm out,0}}\right)
\end{equation}
is the correction coefficient.  When considering this correction factor small, one obtains the scaling relation reported in the second line of \Eq{eq:scaling:Rout}.

In general, the correction coefficient $\mc{C}$ is positive:  the higher the characteristic current is, the stronger the initial shock traversing the liner is and the more compressed the liner becomes.  Therefore, $\Delta R$ increases with the characteristic current, so $\mc{C}$ is positive.

One could attempt to derive a simple physical model to calculate the correction coefficient $\mc{C}$ from first principles.  However, such a model falls beyond the scope of this paper.  Instead, we used the parametric form of the scaling law \eq{eq:app:Rout} and determined $\mc{C}$ by minimizing the difference in simulated bang times between the base load and a scaled load.  Once a satisfying correction coefficient was obtained, we then used the obtained scaling law \eq{eq:scaling:Rout} to generate the initial liner dimensions of the similarity-scaled loads studied here.

%%%%%%%%%%%%%%%%%%%%%%%%%%%%%%%%%%%%%%%%%%%%%%%%%
%%%%%%%%%%%%%%%%%%%%%%%%%%%%%%%%%%%%%%%%%%%%%%%%%
%%%%%%%%%%%%%%%%%%%%%%%%%%%%%%%%%%%%%%%%%%%%%%%%%

%\bibliographystyle{apsrev-title}
%\bibliography{current_scaling,foot}

\begin{thebibliography}{62}
\bibitem[{foo({\natexlab{b}})}]{foot:linerheight}
\bibinfo{note}{Only the axial region of the liner between the top and bottom
  cushions of a MagLIF load implodes and compresses the fuel (see
  \Fig{fig:liners}). We call the axial length between the cushions the
  ``imploding height" or the ``liner height".}
\bibitem[{foo({\natexlab{c}})}]{foot:axial_scaling}
\bibinfo{note}{For the sake of simplicity, we adopted geometric scaling for the
  axial dimensions of the MagLIF load. Other scaling rules can be adopted. For
  example, maintaining the axial dimensions of the cushions as in present-day
  MagLIF might be enough to mitigate the wall instability. This modification in
  the scaling would be more advantageous as the fuel would require less preheat
  energy. Exploring such alternative scaling prescriptions for the axial
  dimensions will be left for future exercises of finding specific MagLIF point
  designs for future pulsed-power generators.}
\bibitem[{foo({\natexlab{f}})}]{foot:preheat}
\bibinfo{note}{In present-day MagLIF experiments on the Z machine, the laser
  pulse length is closer to 5~ns. However, here we choose a 10-ns pulse length
  since a longer pulse length will likely be required for higher preheat
  energies to reduce the laser intensity and the ensuing laser--plasma
  instabilities.}
\bibitem[{foo({\natexlab{g}})}]{foot:timplosion}
\bibinfo{note}{The implosion time is defined as the difference between the
  peak-burn time and the ``zero-current" time, which is calculated via a linear
  fit to the rising portion of the current traces.}
\end{thebibliography}

\end{document}